\documentclass[sigconf,reprint,nonacm]{acmart}

\setcopyright{none}
\renewcommand\footnotetextcopyrightpermission[1]{}

\usepackage{comment}
\usepackage{hyperref}
\usepackage{url}
\usepackage{mathtools}
\usepackage{bussproofs}
\usepackage{pdftexcmds}
\usepackage{stmaryrd}
\usepackage{amsthm}
\usepackage{amssymb}
\usepackage{color}
\usepackage{booktabs}
\usepackage{xspace}
\usepackage[linesnumbered, ruled]{algorithm2e}
\usepackage{soul} 
\usepackage{graphicx}
\DeclareGraphicsExtensions{.jpg,.png,.pdf}
\newcommand{\cC}[0]{\mathcal{C}}

\newcommand{\cH}[0]{\mathcal{H}}

\newcommand{\cU}[0]{\mathcal{U}}
\newcommand{\cX}[0]{\mathcal{X}}
\newcommand{\cY}[0]{\mathcal{Y}}

\newtheorem{theorem}{Theorem}[section]

\newtheorem{example}[theorem]{Example}

\newcommand{\AVC}{AVC\xspace}
\newcommand{\AVCstar}{AVC$^*$\xspace}
\newcommand{\Krimpstar}{OC3$^*$\xspace}
\newcommand{\CompreXstar}{CompreX$^*$\xspace}

\begin{document}

\title{Categorical anomaly detection in heterogeneous data using
  minimum description length clustering}

\author{James Cheney}
\email{jcheney@inf.ed.ac.uk}
\affiliation{%
  \institution{University of Edinburgh}}
\affiliation{\institution{The Alan Turing Institute}}

\author{Xavier Gombau}
\email{xavi.aatt@gmail.com}
\affiliation{%
  \institution{Polytechnic University of Catalonia}
}
\author{Ghita Berrada}
\email{berrada.ghita@gmail.com}
\affiliation{%
  \institution{King's College, London}
}
\author{Sidahmed Benabderrahmane}
\email{sidahmed.benabderrahmane@nyu.edu}
\affiliation{%
  \institution{New York University}
}


\begin{abstract}
  Fast and effective unsupervised anomaly detection algorithms have
  been proposed for categorical data based on the minimum description
  length (MDL) principle.  However, they can be ineffective when
  detecting anomalies in \emph{heterogeneous} datasets representing a
  mixture of different sources, such as security scenarios in which
  system and user processes have distinct behavior patterns. We
  propose a meta-algorithm for enhancing any MDL-based anomaly
  detection model to deal with heterogeneous data by fitting a mixture
  model to the data, via a variant of $k$-means clustering.  Our
  experimental results show that using a discrete mixture model
  provides competitive performance relative to two previous anomaly detection
  algorithms, while mixtures of more sophisticated models yield
  further gains, on both synthetic datasets and realistic datasets from a
  security scenario.
\end{abstract}




\maketitle

\section{Introduction}\label{sec:intro}
A wide variety of anomaly detection techniques have been studied,
considering numerical, categorical, and mixed
data~\cite{anomaly}.  Anomaly detection, or outlier detection,
is based on different strategies for estimating the degree to which
individual data points differ from the norm exhibited by the dataset
as a whole.  This challenge is
usually compounded by the fact that annotated training data indicating
whether data items are normal or abnormal may be unavailable,
unbalanced, or unrepresentative of future observations.
For example, in a security setting, past attacks may not be
representative of future yet-to-be-invented attacks, and attack data
is typically sparse, so training an
accurate binary classifier is likely to either overfit to the past
attacks, or suffer low accuracy against known attacks.  We consider
\emph{unsupervised} anomaly detection.

Most work on unsupervised anomaly detection has focused on continuous, numerical
data.  In this paper, we focus on categorical data, for which several
different techniques have been
studied~\cite{taha19csur}.  One of the most effective
classes of techniques is based on the \emph{minimum description
  length} (MDL) principle~\cite{mdl}.  According
to the MDL principle, we measure how well a model fits the data by how
well it \emph{compresses} the data, plus a cost associated with
representing the model itself.  The idea is to avoid overfitting by
trading off model complexity for accuracy: for example, in the limit,
a model that contains a dictionary listing each possible data value
would compress the data very well, but be penalized highly for model
complexity, since the model contains a copy of all of the different
possible values of data records. In the MDL-based approach to anomaly
detection~\cite{krimp-ad,comprex}, 
we first apply MDL to select a ``good'' model of the data, and then
use the compressed size of each data item as its anomaly score.

Two examples of anomaly detection based on MDL have been studied and
shown to perform well: the OC3 algorithm~\cite{krimp-ad} based on an
itemset mining technique called Krimp~\cite{krimp}, and the CompreX
algorithm~\cite{comprex}.  Broadly speaking, both take a similar
approach: first, a model $H$ of the data that compresses it well is found
using a heuristic search, balancing the model complexity $L(H)$
(number of bits required to compress the model structure/parameters)
against the data complexity $L(X|H)$ (number of bits required to
compress the data given the model).  Once such a model $H$ is found,
we assign to each object $x\in X$ a score corresponding to the object's
compressed size $L(x|H)$ given the selected model.  Intuitively, if
the model accurately characterizes the data as a whole, records that
are representative will compress well, yielding a low anomaly score,
while anomalous or abnormal records should compress poorly.  If this
were not the case, then a more accurate model (compressing the data
better) could be obtained by giving the normal records shorter codes
and anomalous records longer codes.

While effective, these approaches have some limitations.  They work
well for \emph{homogeneous} datasets, for which it is reasonable to
assume that there is a single process that generates the observed
data.  However, the compression models they consider take no account
of the possibility that the data represents a \emph{mixture} of
records generated by different data sources.  If the data is
heterogeneous, there may be further opportunities for compressing it
more effectively, by choosing among several different models instead
of using a single one.  

\begin{figure}[tb]
  \centering
  \includegraphics[scale=0.4]{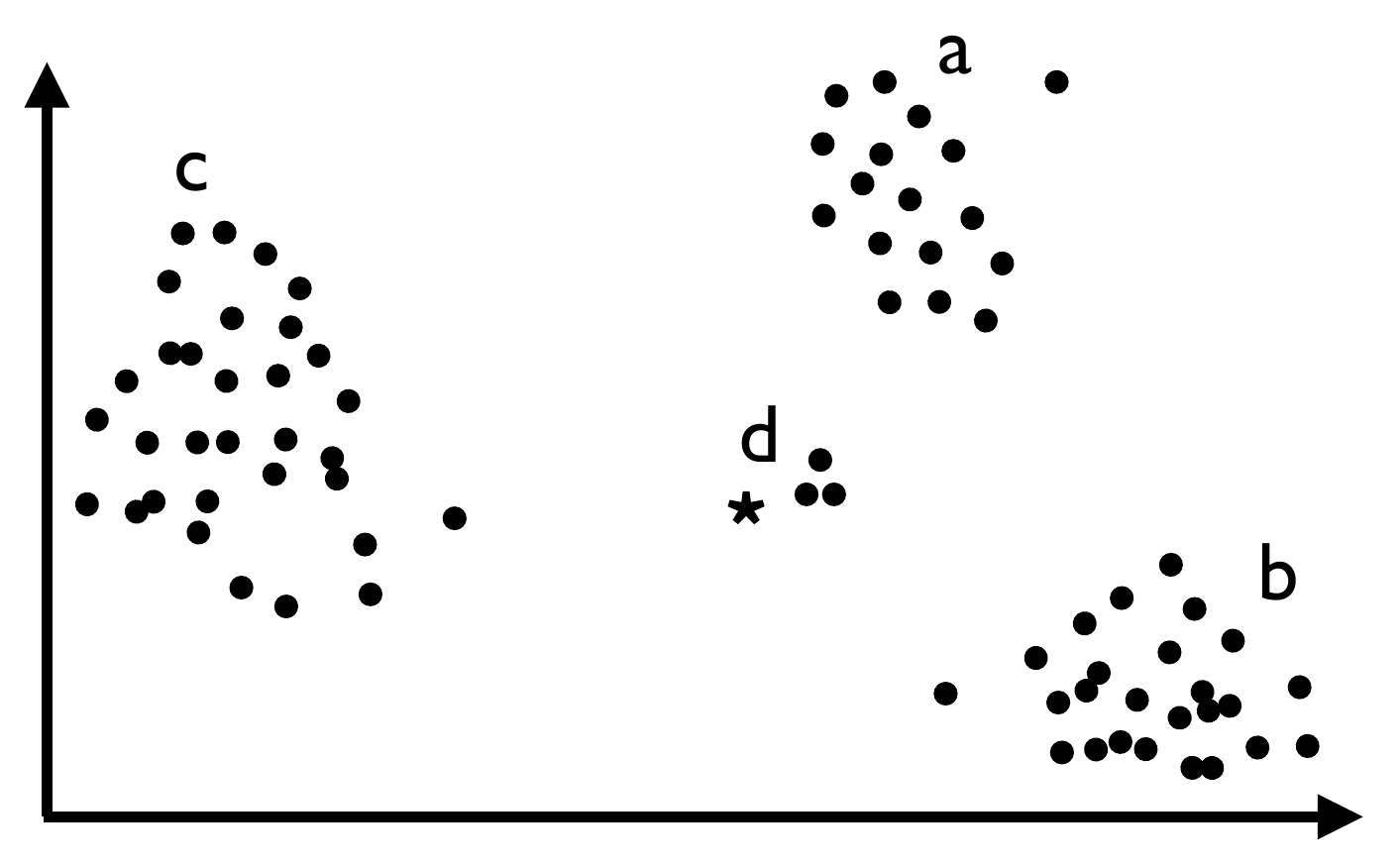}
  \caption{Geometric intuition}\label{fig:cluster}
\begin{tabular}{cc}
(a)&\includegraphics[scale=0.5]{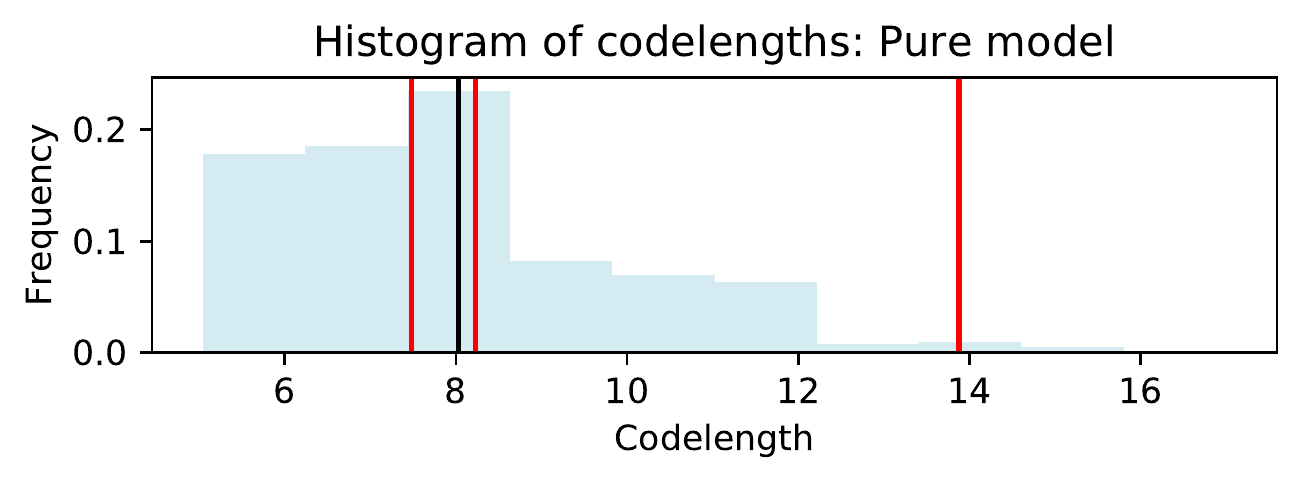}\\
(b)&\includegraphics[scale=0.5]{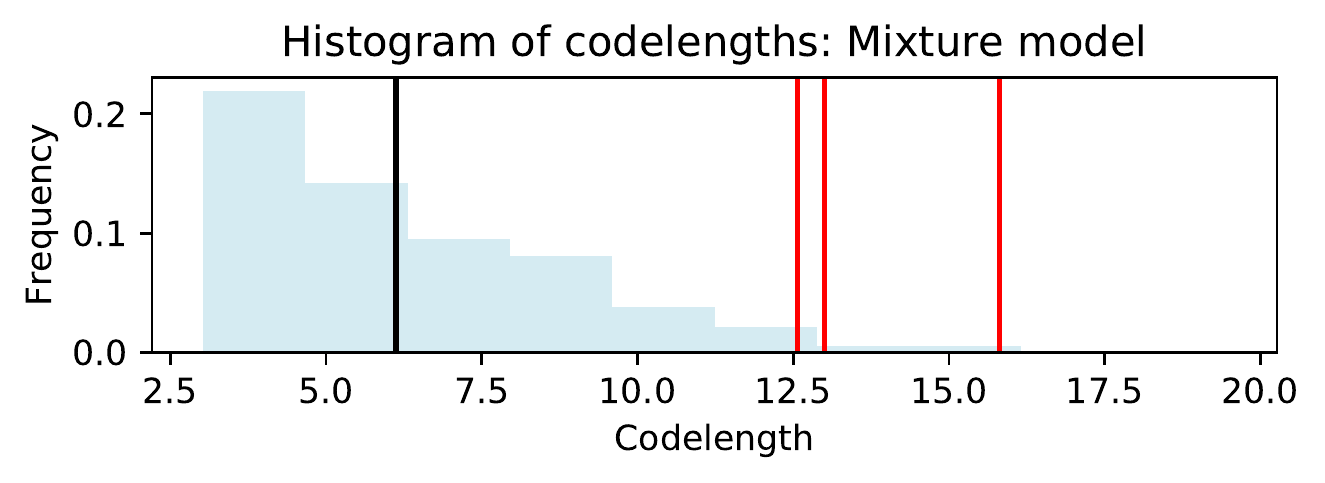}
\end{tabular}
  \caption{Results of (a) pure and (b)  mixture modeling for anomaly detection in
    synthetic data.  Red vertical lines indicate anomalies, while the
    black vertical line indicates the average codelength.}\label{fig:synthetic}
\end{figure}

We illustrate the potential benefits of mixture modeling using
geometric intuition.  Figure~\ref{fig:cluster} illustrates some data
points in a two-dimensional space, with three large clusters (a,b,c),
and a small cluster (d).  If we consider a single model of the data,
the ``center'' of the dataset (illustrated with a star) is closest to
the points (d), while points (a,b,c) are approximately equidistant.
Thus, if we used distance from the center as an anomaly score, it
would be difficult to distinguish the anomalous cluster from the three
main clusters.  If, on the other hand, we recognize that most of the
data fall into three main clusters, then we would see that the points
(d) are not close to any of these clusters, even though they are close
to the average  behavior of the dataset as a whole.

We have emulated this situation in a categorical setting by generating
synthetic data with 1000 samples drawn from three discrete
distributions corresponding to the large clusters (a,b,c), and seeding
just three data points for the anomalous cluster (d), whose
distribution is the average of the three clusters.  The results are
shown in Figure~\ref{fig:synthetic}.  In this synthetic
dataset, fitting a single MDL model to the data (as shown in Figure
2a) results in the seeded anomalies being detected at ranks 35, 1315
and 1655, and two of the anomalies have scores close to the
average.  On the other hand, fitting a mixture model yields
compression savings of over 30\%, indicating the mixture model is much
more likely to be correct according to the MDL principle, and ranks
the seeded anomalies at positions 17, 42 and 109.  (These results were
generated using the AVC and \AVCstar algorithms described later in the
paper.)  Note that this approach can be viewed (at least in spirit) as
a form of \emph{outlier ensembling}~\cite{aggarwal17}; however, to the best of
our knowledge there has been no previous work on ensembling anomaly
detectors based on minimum description length.

In this paper, we propose an MDL-based anomaly detection technique
that exploits this opportunity to detect and exploit heterogeneities
in datasets to improve anomaly detection.  Our approach is partly
inspired by work using Krimp for
clustering~\cite{identifying-the-components}.  This work proposed both
top-down and bottom-up clustering strategies, using Krimp's
compression metric to assess the quality of clusters, and comparing
the quality of clusterings obtained for different values of $k$.  They
found up to 40\% improvement in compressed size.  The top-down
algorithm they presented is similar to the k-means algorithm for
clustering numerical data~\cite{bishop06}, and is similar to other MDL-based
clustering algorithms~\cite{kontkanen06mdl}.
They observed this approach generalizes to any MDL-based technique but
did not explore this insight or consider applications to anomaly detection.

Based on this observation, we consider the value of MDL-based
clustering as a basis for anomaly detection.  Essentially, the idea is
to first identify a clustering for the data (and, in the process,
assign each data item to the most appropriate cluster).  Next, a score
is assigned to each data item based on combining the cost of
compressing its cluster number, together with the cost of compressing
the data itself using the corresponding model.  Both components would
be required to decompress the data, since, without the cluster number,
the receiver would not know which basic model to use to decompress the
data.

In this paper, we consider several instances of this idea.  Our main
contribution is to demonstrate that clustering techniques can improve
the anomaly detection performance of a variety of MDL-based models.
After introducing notation and a framework for MDL
algorithms~(Section~\ref{sec:background}), we consider a naive
MDL-based anomaly detection technique called AVC, which is a slight
variant of the AVF algorithm~\cite{avf} in which the data is fit to a
product of independent Bernoulli distributions.  We consider AVC,
Krimp/OC3, and CompreX as instances of the MDL framework, and
describe a meta-algorithm that uses a clustering strategy to fit a
mixture model to the data  (Section ~\ref{sec:mdl}).  We then present
experimental evaluation (Section~\ref{sec:evaluation})
using common datasets as well as data from a realistic security
setting which demonstrates that clustering can improve anomaly
detection performance (measured using AUC or nDCG score)
significantly, while also imposing higher computational cost.
Interestingly, the benefits of clustering are much more significant
for AVC compared to Krimp/OC3 or CompreX, illustrating that the more
sophisticated compression techniques used by the latter are already
flexible enough to deal with heterogeneous data.  In some cases,
clustering using mixtures of AVC models actually outperforms (mixtures
of) Krimp or CompreX models, while also being faster.

\section{Background}\label{sec:background}
We assume
familiarity with basics of information theory and the MDL
principle~\cite{mdl,coverthomas}. We consider the following framework for selecting and fitting models
to data according to the MDL principle.  (We employ what
Gr\"unwald~\cite{mdl} calls ``two part crude MDL'', which is
well-behaved when large amounts of data are available.)  We consider an MDL scenario to be
specified as follows:
\begin{itemize}
\item A space of \emph{observations} $X$
\item A space of \emph{hypotheses} $H \in \cH$
\item A global hypothesis codelength  function $L : \cH \to \mathbb{R}$
  satisfying the Kraft inequality $\sum_{H \in \cH} 2^{-L(H)} \leq 1$.
\item A function $L(-|-) : X \times \cH \to \mathbb{R}$, such that for each
  hypothesis $H$ the  associated codelength
  function $x \mapsto L(x|H)$ takes a data item $x\in X$ to its compressed
  size, again satisfying the Kraft inequality $\sum_{x\in X}2^{-L(x|H)}
  \leq 1$.
\end{itemize}
Given a MDL scenario $(X,\cH,L)$, a \emph{learner} is an algorithm
$F : X^n \to \cH$ that attempts to find $H$ minimizing
$L(H) + \sum_{i=1}^n L(x_i|H)$.  It is not required that a learner
finds a global optimum $H_{\min}$; we also allow for $F$ to be
nondeterministic or randomized.  Familiar probability distributions,
with their associated parameter estimation techniques, provide a ready
source of MDL-based learning techniques, but our framework also
permits viewing any family of compression algorithms (indexed by
hypotheses corresponding to parameterizations, code tables, etc.) as
an MDL-based learning technique.

For the purpose of this paper we assume the size of the dataset under
consideration is a fixed constant $n$.

\paragraph{Uniform processes} As a simple case we consider the trivial
MDL scenario $\cU_X$ corresponding to a uniform probability distribution over
a finite set $X$.  In this case, there is only one hypothesis, $\star$,
representing the uniform distribution, and the hypothesis codelength
function is $L(\star) = 0$, while the observation codelength function
is $L(c_i|\star) = -\log (\frac{1}{|X|}) = \log |X|$.

\paragraph{Bernoulli processes}
We consider Bernoulli processes, generating 0,1 values, specified by a
probability $p$ of generating a 1 and $1-p$ of generating a 0.  The
hypotheses we might seek to learn about such a process are the
probabilities $p$; to avoid degeneracy we require $0 < p < 1$.  To
avoid having to specify real numbers infinitely precisely, we consider
precision to $\lfloor\log n\rfloor$ bits, where $n$ is the number of data
items under consideration.  We represent a hypothesis as $m$
interpreted as a rational number $m/n$ where $0 \leq m \leq n$.  The
codelength of such a hypothesis is $L(m) = \lfloor\log n\rfloor$, we
encode $m$ using $\lfloor\log n\rfloor$ bits.

Given a sequence of $n$ independent 0/1 values we may estimate
$p = n_{[1]}/n$ where $n_{[1]}$ is the number of 1's occurring in the
sequence.  This choice would be optimal, if we wish to compress the
sequence minimizing the sum $\sum_{i=1}^n L(x^i|H)$ so that there is no need to deal with data values that were not observed in the input.
However, this leads to degeneracy in the case where some value
never appears in the input, which means that future observations of
unseen values have probability 0, and hence infinite codelength.  To
avoid this problem, we generally apply a so-called \emph{Laplace
  correction} to the counts, by taking $p = (n_{[1]}+1)/(n+2)$ to
ensure that both outcomes have nonzero probability.

Thus, the
codelength of a 0 or 1 value $x$ is $-\log p$ if $x = 1$ and
$-\log(1-p)$ otherwise; we may write this concisely as
$L(x|H) = -[x=1]\log p - [x=0]\log(1-p)$, where $p = (m+1)/(n+2)$ is derived
from $H = (m,n)$ as above.  (We write $[\phi]$ for $0$ if $\phi$ is
false and $1$ if $\phi$ is true.)
Finally, we take the learning function $F : \{0,1\}^* \to \cH$ to be 
\[F(x^1,\ldots,x^n) = \sum_{i=1}^n x^i\]

\paragraph{Categorical processes}
We can generalize the above discussion of two-valued Bernoulli
processes to $k$-valued categorical processes $\cC_k$.  A categorical
distribution over $k$ outcomes $1,\ldots,k$ is specified by the $k$
probabilities $p_1,\ldots,p_k$ of the different outcomes, up to
precision $\lfloor\log n\rfloor$; since one of them is redundant, the data of a
hypothesis is given by $(m_1,\ldots,m_{k-1})$ and the associated
codelength $L(H) = (k-1)\lfloor\log n\rfloor$.  From this, we may
extract $p_1,\ldots,p_k$ by taking $p_i = (m_i+1)/(n+k)$ where $i < k$
and $p_k = 1-\sum_{i<k}p_i$, again applying Laplace correction to
avoid degeneracy.  Furthermore we may encode a given outcome $x$
according to hypothesis $H$ as $\sum_{j=1}^k -[x=j]\log p_j$.

Finally the learning function that identifies the best hypothesis for
the data is:
\[F(x^1,\ldots,x^n) = (\sum_{i-=1}^n [x^i=1],\ldots,\sum_{i=1}^n[x^i=k-1])\]

\paragraph{Independent products}

Given two MDL scenarios $\cX = (X,\cH_1,L_1)$ and $\cY=(Y,\cH_2,L_2)$, we can
combine them independently to form a scenario over pairs
$(x,y) \in X\times Y$ by taking a product $\cX\times \cY$.  The
hypotheses $\cH = \cH_1\times \cH_2$ for the product are pairs of
hypotheses for $\cX$ and $\cY$.  The hypothesis codelength function is
defined by taking sums of codelengths:
$L(H_1,H_2) =L_1(H_1) + L_2(H_2)$.  The data codelength function is
also defined by taking sums, using the respective hypotheses:
$L((x,y)|(H_1,H_2)) = L_1(x|H_1) + L_2(y|H_2)$.  Finally, given
learning functions $F_1,F_2$ for $\cX$ and $\cY$ respectively, the
learner function $F$ is defined as:
\[F((x^1,y^1),\ldots,(x^n,y^n)) = (F_1(x^1,\ldots,x^n),F_2(y^1,\ldots,y^n))\]

Likewise, we can also consider $n$-ary products $\prod_{i=1}^n \cX_i =
\cX_1\times \cdots
\cX_m$ whose behavior is determined by iterating binary products.  

\section{MDL-based anomaly detection}\label{sec:mdl}
We now consider several instances of the above framework and their use
for anomaly detection.  In each case, we follow the same recipe as
proposed in previous work~\cite{krimp-ad,comprex}: first induce a good
model of the data according to the MDL principle, that compresses the
dataset as a whole; then assign each element an anomaly score
corresponding to its compressed size using the model.  We can then
inspect the highest-scoring data items as being the most anomalous.

The first instance of this framework, called \emph{Attribute Value
  Compression} (AVC), is a slight variation of a known anomaly
detection algorithm called \emph{Attribute Value Frequency}
(AVF)~\cite{avf}, but is based on optimal compression of each
attribute independently.  Thus, it is limited in that it cannot
exploit interdependencies among attributes.  Next, we recast the
Krimp/OC3 and CompreX algorithms as instances of the above framework,
though we abstract over the details.  Finally, we describe a generic
strategy (i.e.\ a meta-algorithm) for learning mixtures of MDL
hypotheses, given a basic learner such as AVC, Krimp or CompreX.
Given a learner $F$ for component models, the meta-algorithm produces
a learner for $k$-mixture models $F^*$, using the MDL principle to
find a suitable value for $k$.
\subsection{Basic models}
\paragraph{Attribute Value Coding}
Given a binary or categorical dataset with $m$ attributes
$\cX_1,\ldots,\cX_m$, each corresponding to a Bernoulli or categorical
process, we define the \emph{Attribute Value Coding} algorithm as
the learner induced by taking the product of the optimal learners for
the components $\cX_1,\ldots,\cX_m$.  Concretely, $F_{AVC}$ calculates
the frequencies of the values of each attribute independently and
$L_{AVC}(x_1,\ldots,x_m| H)$ then encodes each attribute $x_i$ with
codelength $-\log p_i$ where $p_i$ is the probability of attribute
$i$ having value $x_i$.  For example, when the attributes are all
binary, the probabilities first calculated as $p_i =
\frac{1}{n}\sum_{j=1}^n x^j_i$ and the score of each element $x$ is
$L(x|H) = \sum_{j=1}^m -([x_j=1] \log p_j + [x_j=0] \log (1-p_j))$.

\begin{example}
    Here, we introduce a small (extremely simplistic) running example to illustrate AVC.
  Suppose we have the following dataset:
    {
\[\begin{array}{c|ccc}
id & \texttt{abc.com} & \texttt{xyz.com} & \texttt{evil.com}\\\hline
P_{17} & 1 & 1 & 0\\
P_{42} & 1 & 1 & 0\\
P_{1337} & 0 & 0 & 1\\
P_{007} & 1 & 1 & 1 
\end{array}\]
}
  where $P_{17}, P_{42}, P_{1337}, P_{007}$ are four distinct processes and $abc.com$, $xyz.com$ and $evil.com$, three attributes
  corresponding to network addresses accessed by the processes.
  
  $P_{17}$ and $P_{42}$ access both \texttt{abc.com} and \texttt{xyz.com} and are both processes with innocuous activity while $P_{1337}$ and $P_{007}$ are malicious processes ($P_{1337}$ a naive attacker that only accesses \texttt{evil.com}  and $P_{007}$ a more sophisticated attacker that accesses all three addresses in order to attempt to camouflage its behavior).  
  
  To calculate the AVC score of each of the processes and determine which exhibit abnormal behavior, we first compute the frequencies of occurrence ($c_i$, with $i \in \{abc.com,xyz.com,evil.com\}$) of each of the three dataset attributes followed by the probability $p_i$ of each of the attributes having value $x_i$ (where $x_i$ is either 0 or 1):
  {
  \begin{eqnarray*}
   &c_{\texttt{abc.com}}=c_{\texttt{xyz.com}} = 3\quad\textrm{and}\quad c_{\texttt{evil.com}} = 2\\
   &p_{\texttt{abc.com}=1}=p_{\texttt{xyz.com}=1} = \frac{3}{4}\quad\textrm{and}\quad p_{\texttt{evil.com}=1} = \frac{1}{2}
  \end{eqnarray*}
  }
(the probabilities can be estimated simply by taking $p_i =
\tfrac{c_i}{n}$ where $n$ is the number of data points i.e. processes here)

  The AVC scores are then simply calculated as follows:
  {
  \begin{eqnarray*}
    \AVC(P_{17})  &=& -\tfrac{1}{3}(\log\tfrac{3}{4} + \log\tfrac{3}{4} + \log\tfrac{1}{2}) \approx 0.61\\
    \AVC(P_{42})  &=& -\tfrac{1}{3}(\log\tfrac{3}{4} + \log\tfrac{3}{4} +
                     \log\tfrac{1}{2}) \approx 0.61\\
    \AVC(P_{1337})  &=&  -\tfrac{1}{3}(\log\tfrac{1}{4} + \log\tfrac{1}{4} +
                     \log\tfrac{1}{2}) \approx 1.66\\ 
    \AVC(P_{007})  &=& -\tfrac{1}{3}(\log\tfrac{3}{4} + \log\tfrac{3}{4} + \log\tfrac{1}{2}) \approx 0.61
  \end{eqnarray*}
  }
\end{example}
We term this approach ``Attribute Value Coding'' because it is
very similar to a previously-defined anomaly detection technique
called ``Attribute Value Frequency'' (AVF)~\cite{avf}.  The main
difference is that, in AVF, we sum the \emph{probabilities} of each
attribute attaining its observed value, not the
\emph{log-probabilities} (i.e. codelengths); hence, AVF does not have a direct reading
as an MDL technique, while AVC does.  Because AVF scores do not
correspond to compressed sizes, it would be meaningless to attempt to
use them as a component in a larger MDL-based compression strategy
such as the one we propose based on clustering.  

\paragraph{Krimp}

We will also consider the Krimp algorithm as an MDL scenario and
learner.  In Krimp, a (binary) dataset is compressed by identifying
certain subsets of frequently co-occurring attributes.  Concretely, in
Krimp, the dataset is to be represented by a \emph{code table} $CT$
which lists the possible subsets and their codelengths, and then each
data item is represented by a set of codewords called its
\emph{cover}, so that the length
$L(x|CT) = \sum_{y \in \mathit{cover}_{CT}(x)} \mathit{length}(y)$.

In Krimp, a candidate collection of itemsets is mined from the data
using standard techniques.  The mined itemsets are considered as
candidate entries to a code table; those that are useful in improving
compression are selected.  Krimp performs a heuristic search to try to
find a code table that minimizes the compressed size of the data.
Different pruning strategies are used to remove candidates which are
less effective.  The exact details of the search and pruning
algorithms do not play an important role here as we will use them as a
black box.

However, what is important is that we can assign a cost $L(CT)$ to the
code table itself (that is, the number of bits necessary to record or
communicate it) and we can assign a cost $L(x|CT)$ to each data item
given a code table (that is, the number of bits necessary to
communicate a given data item, using a certain code table). These
codelengths satisfy the Kraft inequality.

In the MDL scenario corresponding to Krimp, the hypotheses are Krimp's
code tables $CT$, and the codelength functions $L(CT), L(-|CT)$ are as
defined in previous papers~\cite{krimp,krimp-ad}.  We write
$F_{\mathrm{Krimp}}$ for the Krimp algorithm itself, which selects
among the (huge number of) potential code tables one which performs
well in balancing the hypothesis codelength against the encoded size
of the data.  

\paragraph{CompreX}

We will also describe how to model the CompreX algorithm and its MDL
scenario.  In CompreX, as in Krimp, code tables are used.  However, in
CompreX, instead of using a single global code table, the input
attributes are \emph{partitioned} and one code table is assigned to
encode the attributes in each component of the partition.  Thus, the
hypotheses $H = (P,CT_1,\ldots,CT_p)$ consist of a partition
$P = \{P_1,\ldots,P_p\}$ of the attributes, along with one code table
for each partition.  A hypothesis codelength function $L(H)$ is
described in the paper~\cite{comprex} using Krimp's $L(CT)$ together
with an encoding of partitions, and the codelength function for data
elements given a hypothesis is derived by adding together the
codelengths of the attributes in each part:
\[L(x|H) = \sum_{i = 1}^p L(\pi_{P_i}(x) | CT_i)\]
where $\pi_{P_i}(x)$ is the projection of the attributes of $x$ to the
subset $P_i$.  Note that once the partition is given, this is
essentially a product of Krimp MDL scenarios.

Unlike Krimp, CompreX follows a bottom-up strategy for synthesizing
code tables: initially the partition consists of singleton attributes
only and the associated code tables are trivial.  Using information
gain as a heuristic, CompreX greedily merges partition elements and
combines their code tables.  Itemset mining is not directly performed;
nevertheless, CompreX was found to obtain good compression compared to
Krimp, indicating that many datasets may contain subsets of
highly-correlated attributes for which CompreX's partitioning strategy
works well.  We write $F_{\mathrm{CompreX}}$ for CompreX considered as
an MDL learner algorithm, relative to an appropriate MDL scenario.
\subsection{Mixture models}
We assume given a basic MDL scenario $\cX = (X,\cH,L)$ and learner $F$
for fitting hypotheses to the data.  As mentioned
by~\cite{identifying-the-components}, any such technique can be used
as a component in a $k$-means-style clustering technique, which can
again be justified by the MDL principle.  In this section, we spell
out the details in a way that is independent of the choice of $\cX$
and $F$.

Given the MDL scenario $\cX = (X,\cH,L)$, we can construct a new scenario $\cX^*$
called the $\cX$-mixture model as follows:
\begin{itemize}
\item The observations $X$ are those of $\cX$.
\item The hypotheses correspond to $K$-mixture models of hypotheses
  $\cH$, for all positive $K$.  These are tuples $(K,H,
  \vec{H_i})$, where $H$ is a hypothesis for the $K$ possible
 components specified by $\cC_k = (\{1,\ldots,K\},\cH_K,L_K)$, the $K$-valued categorical
  process model and the $K$ hypotheses $H_1,\ldots,H_K\in \cH$
characterize each component.
\item Define the encoding for a hypothesis $(K,H,\vec{H_i})$ as
  $L^*(K,H,\vec{H_i}) = \lfloor \log n \rfloor + L_K(H) +
  \sum_{i=1}^kL(H_i)$.
  That is, we encode $K$ (which takes at most $\lfloor \log n \rfloor$
  bits), then the hypothesis for the class labels $H$ (using the
  codelength function from $\cC_K$), and finally the $K$ hypotheses
  for the data for each class (using the hypothesis codelength
  function from $\cX$).
\item Define the encoding for each data value $x$ as $L^*(x|(K,H,\vec{H_i})) =
  \min_{i\in \{1,\ldots,K\}} L_K(i|H) + L(x|H_i)$.
\end{itemize}
We write $\cX^*$ for the MDL scenario specified above, and call it the
\emph{mixture of $\cX$ models}.

Intuitively, this scenario corresponds to the following
(nondeterministic) compression
algorithm:  we guess $K$, a distribution over $K$ class labels, and $K$ hypotheses corresponding to the $K$
components existing in the data.  We encode this information and
transmit it to the receiver.  Subsequently, each data value $x$ can then be
transmitted by first encoding the class label $i$  for $x$ (using the
hypothesis $H$ describing the distribution of class labels), then
transmitting $x$ itself using the hypothesis $H_i$.  

This nondeterministic algorithm suggests an optimal (but infeasible)
compression algorithm: given data $x \in X^*$, find a mixture model
hypothesis that yields the optimal codelength $L(H^*) + L(x|H^*)$
given the above scenario.  Of course, this naive approach is infeasible since, even
if we know an efficient optimal learner for the components $\cX$,
finding the optimal mixture model parameters (equivalently, the
optimal clustering minimizing the codelength) might require
considering all of the possible partitions.  The number of
partitions of $n$ is the Bell number $B_n$
which grows very rapidly (e.g. $B_{20} > $ 50 trillion).  

However, just as for conventional clustering, an iterative greedy approach can be
effective, following a similar strategy to the classic $k$-means
clustering algorithm.  We outline such a strategy below, which is
largely the same as in~\cite{identifying-the-components}; the main
difference is that we use an optimal code for the class labels,
instead of ignoring the codelength of the class labels as they do.

The mixture model fitting process is a variant of the $K$-means
clustering algorithm, but using MDL hypotheses to represent the
clusters, and with codelength assigned to a point by a given cluster
playing a role analogous to the distance metric in $K$-means.  To find
the right $k$, we start with $k =1$ and increase it until we have
found a local minimum.  (In practice, we typically detect when
increasing $k$ yields diminishing returns, and stop early, since trying all $k$ up to
$n$ would be prohibitively expensive.)
\begin{algorithm}
  \KwIn{a dataset $\vec{x}$}
  \KwOut{An $\cX^*$ hypothesis $(K,H,\vec{H_i})$ }
  $k=1$\;
  \While{$k\leq n$}{
    Randomly
assign each $x^i$ to one of $k$ classes. Let $y^i$ be the initial
class of each $x^i$\;
    \Repeat{a local minimum $c_k$ is reached}{
      Run $H_j := F_\cX(\{x^i \mid y^i = j\})$ to
calculate the hypothesis for each class\;
   Re-assign each $x_i$ to
the class $j$ minimizing $L(x^i | H_j)$, setting $y^i$ to $j$\;
Evaluate $c := L^*(K,H,\vec{H_i}) + \sum_{i=1}^{n}L^*(x^i |
(K,H,\vec{H_i}) $, the cost of the current hypothesis\;
    }
    Set $k := k+1$\;
  }
  Return the hypothesis $(k,H,\vec{H_i})$ achieving minimal $c_k$.
  \caption{Mixture model fitting algorithm}
  \label{alg:mixtureModelFitting}
\end{algorithm}
%
Similarly to $k$-means, in each iteration, we alternate between estimating the component
models based on the current candidate clustering (line 5 in Algorithm \ref{alg:mixtureModelFitting}), and reassigning points
to clusters (line 6 in Algorithm \ref{alg:mixtureModelFitting}).  However, instead of taking the ``mean'' of a set of
points, which is meaningless for categorical data, we represent a set
of points using a hypothesis and we calculate the
``distance'' of each point as its (idealized) compressed size, were it
to be compressed using that hypothesis.  The hypotheses may be simple
AVC models, Krimp code tables, CompreX hypotheses, or
those of any other MDL scenario.
Likewise, we might consider different models for the class labels; we
have in mind compressing the class labels optimally according to the
observed distribution of classes, but we could also fix the uniform
distribution $\cU_X$ (which would give the same behavior as the
clustering algorithm of van Leeuwen et
al.~\cite{identifying-the-components}), where the cost of encoding the
class label is ignored; this has the same effect as assuming the class
labels all have the same codelength.)

According to line 8 of Algorithm~\ref{alg:mixtureModelFitting}, we
repeat this process until convergence to a local minimum. Several
techniques for detecting convergence are possible; we fix some small
$\epsilon > 0$ and iterate until the total compressed size fails to
improve by more than $\epsilon$ times the previous size.  We may also
conduct several trials with different random initializations and take
the result that minimizes the compressed size.

Now, to find the best $k$, we consider a range of $k$ values and fit a
model with $k$ components for each $k$. 
Suppose we have constructed models
$\mathcal{M}_1,\ldots,\mathcal{M}_n$ for all possible $k$ values
between 1 and $n$.  We choose $k$ to be the one for which
$L(\mathcal{M}_k) + L(X|\mathcal{M})$ is minimized.  Of course, it
would likely be wasteful to fit models again and again so, by fixing some
$\epsilon$, we may terminate the process early if increasing $k$ fails
to result in a better fit.  In practice, we usually consider $k$
values up to some relatively small number, since $n$ is usually
very large.


\subsection{Anomaly detection}


Let $H$ be the best hypothesis
found by the above procedure for some MDL scenario $\cX$.  To perform anomaly scoring
we simply assign each $x$ its codelength $L(x|H)$. Because we have
required codelength functions to be nondegenerate (i.e. satisfying the
Kraft inequality for each hypothesis), $L(x|H)$ is well-defined
and finite, even if we consider records $x$ that were not present in
the original dataset.  Since codelength functions satisfying the Kraft
inequality correspond to (sub)probability distributions, the records
for which the codelength $L(x|H)$ are largest are precisely those
whose conditional probability given $H$ are smallest.  

\section{Evaluation}\label{sec:evaluation}
\subsection{Datasets}
We consider several public datasets collected for evaluation of categorical
anomaly detection by Pang~\cite{pang16data}.  Their characteristics
are summarized in the first few columns of Table~\ref{tab:q1results}.
The datasets range in size
up to 2k records and between 22--114 attributes, with between 27 and
60 anomalies.  We transformed all
datasets to use binary encodings of multivalued attributes to ensure
compatibility with Krimp. Most of these datasets are derived from
standard classification datasets by choosing one class to be the
normal class and selecting a few examples of another class to be the
anomalies.

We also consider several datasets derived from the DARPA Transparent
Computing program, consisting of data about operating system processes
in a system under attack by an advanced persistent threat (APT) (see
Table~\ref{tab:q1results} lower left for dataset statistics).
These categorical datasets are derived from much larger raw provenance trace
datasets as described by Berrada et al.~\cite{berrada20fgcs}; we
consider only their \emph{ProcessEvent} datasets in which each record
describes the behavior of an operating system process, and the
attributes indicate whether the process ever performed a particular
kind of action (read, write, forking another process, etc.)
These datasets contain a mix of
system and user-level processes describing all of the
activity in an operating system.  These datasets are drawn
from a realistic security application of anomaly detection, and are
mostly larger and with a smaller percentage of anomalies, and
because of the heterogeneity of the underlying OS processes being
recorded we believe they form a more compelling test for our approach
than the generic datasets.  The datasets include two security
evaluation scenarios in which computers running different operating
systems were attacked by simulated APT intruders, usually leading to a
very small percentage of attack processes which we would like to
detect as anomalous.

\subsection{Experimental Results}
\paragraph{Experimental setup}
We implemented AVC and \AVCstar in Python, using libraries for linear
algebra to perform the iterative clustering steps efficiently.  We
also implemented Python scripts that run adapted versions of Krimp/OC3 and
CompreX to perform their clustering variants \Krimpstar and
\CompreXstar.  We made minor changes to publicly available code for
Krimp\footnote{\url{http://eda.mmci.uni-saarland.de/prj/krimp/}}
and
CompreX\footnote{\url{http://eda.mmci.uni-saarland.de/prj/comprex/}}
to enable this.  We used the default settings for Krimp, and
considered all closed itemsets.  For CompreX, we used the default
behavior in the publicly available Matlab implementation.

All experiments were run on an HP Elitedesk 800 with Intel i5-6500 CPU
and 32GB RAM running Scientific Linux 7.

\paragraph{Evaluation metrics}

Following most work on anomaly detection, we report the \emph{AUC
  score} resulting from the ranking induced by the anomaly scores,
i.e. the area under the \emph{receiver operator characteristic} curve,
which summarizes the effectiveness of anomaly detection at all
possible thresholds.  For large datasets with very sparse anomalies,
the AUC score is not very informative because even a very high score
such as 0.999 can correspond to all of the anomalies being found in
the top 0.1\% of records, but this could still be useless for actually
finding anomalies if the dataset has millions of records.  As a
complement to AUC score, we follow~\cite{berrada20fgcs,berrada19tapp}
in also reporting the \emph{normalized discounted cumulative gain} of
the rankings, which is widely used in information retrieval to assess
the results of search algorithms and assigns proportionately greater
weight to rankings that return relevant results (in this case,
anomalies) close to the top. Both nDCG and AUC scores are between 0
and 1, with 1 representing the best possible result.  Their
calculations are otherwise standard and we refer
to~\cite{berrada20fgcs} or other sources for definitions and baseline
results using a variety of categorical anomaly detection algorithms.

\begin{table*}[tb]
\centering
\caption{(Left) Dataset characteristics ($N$=number of records,
  $M$=number of attributes, \%Anomaly = percentage of anomalies). \\
(Right) AUC and nDCG scores for Krimp, CompreX, AVC and \AVCstar.
  The best score of each kind is highlighted in bold.}\label{tab:q1results}
\begin{tabular}{lccccccccccc}
\toprule
&&&& \multicolumn{2}{c}{ Krimp} & \multicolumn{2}{c}{CompreX} & \multicolumn{2}{c}{AVC} &\multicolumn{2}{c}{\AVCstar}\\
\cmidrule{5-12}
Dataset & $N$ & $M$ & \%Anomaly & AUC & nDCG & AUC & nDCG & AUC & nDCG & AUC & nDCG\\
\midrule
AID362 & 4,279 & 114 & 1.4\% & 0.582 & 0.409 & \textbf{0.675} & 0.423 & 0.644 & 0.420 & 0.674 &
                                                                  \textbf{0.433}\\
Bank & 41,188 & 52 &  11\%& 0.625 & 0.814 & \textbf{0.639} & \textbf{0.823} & 0.593 & 0.808 & 0.608
                           & 0.810 \\
Chess (KRK) & 28,056 & 40 & 0.01\%& 0.321 & 0.220 & 0.622 & \textbf{0.263} & 0.645 & 0.244 &
                                                                 \textbf{0.673} & 0.254 \\
CMC & 1,473 & 22 & 2.7\%& 0.559 & 0.402 & 0.580 & \textbf{0.458} & 0.589 & 0.474 &
                                                               \textbf{0.600} & 0.414\\
Probe & 64,759 & 82 &  6.4\% & 0.938 & 0.925 & 0.937 & 0.915 & \textbf{0.951} & \textbf{0.961} &
                                                                  0.937
                           & 0.912 \\
SolarFlare & 1,066 & 41 & 4\% & 0.792 & \textbf{0.593} & \textbf{0.837} & 0.588 & 0.826 & \textbf{0.593} & 0.783 &
                                                                     0.545
  \\
\midrule
 Windows S1 &17,569 & 22 & 0.04\%& 0.992 & 0.302 & \textbf{0.996} & 0.602 & 0.984 & 0.618 &\textbf{0.996} &
                                                                    \textbf{0.675} 
\\
BSD S1 & 76,903& 29 & 0.02\%&\textbf{0.976 }& 0.436 & \textbf{0.976} & \textbf{0.542} & 0.882 & 0.525 & 0.975 & 0.516 \\
Linux S1  & 247,160& 24 & 0.01\%& \textbf{0.887} & 0.340 & \textbf{0.887} & 0.299 & 0.821 & 0.264 & \textbf{0.887} &
                                                                    \textbf{0.407} 
\\
Android S1 & 102& 21& 8.8\%& 0.754 & 0.740 & 0.731 & 0.821 & 0.826 & 0.848 &
                                                            \textbf{0.860} & \textbf{0.861}\\
\hline
 Windows S2  & 11,151 & 30 &0.07\%& 0.857 & \textbf{0.242} & 0.856 & 0.223 & 0.808 & 0.230 & \textbf{0.881} &
                                                                       0.240 
\\
BSD S2 & 224,624& 31& 0.004\%& \textbf{0.936} & \textbf{0.249} & 0.904 & 0.211 & 0.873 &
                                                                   0.191 & 0.917 & 0.186
\\
Linux S2  & 282,087& 25& 0.01\%& 0.873 & 0.387 & \textbf{0.873} & \textbf{0.469} & 0.8240 & 0.306 & 0.856 &
                                                                     0.358 
\\
Android S2 & 12,106& 27& 0.1\%& 0.884 & 0.328 & \textbf{0.930} & \textbf{0.780} & 0.906 & 0.305 & 0.907 &
                                                                      0.629 
\\
\bottomrule
\end{tabular}
\end{table*}

\paragraph{Research questions and experiments}

We ran experiments intended to assess the following research questions:
\begin{enumerate}
\item Q1: Can clustering using simple AVC models yield anomaly detection performance competitive with Krimp or
  CompreX?
\item Q2: Can clustering increase the anomaly detection
  performance of Krimp or CompreX?
\item Q3: Is the performance overhead of clustering acceptable?
\end{enumerate}

To assess Q1, we evaluated the anomaly detection performance of AVC, Krimp,
CompreX, and \AVCstar on the different datasets. The results are
summarized in Table~\ref{tab:q1results}.  In the case of AVC, Krimp
and CompreX, the reported result is the result of one run since the
result is deterministic.  For \AVCstar, we ran 10 runs for each dataset
with different random initializations, since $k$-means algorithms are
sensitive to initial conditions, and we report the median result from
the 10 runs for the $k$ value yielding the smallest compressed sizes.

\begin{figure*}[tb]
\begin{tabular}{cccc}
Bank (\AVCstar) & Bank (\Krimpstar) &Probe (\AVCstar) & Probe (\Krimpstar) \\

\includegraphics[scale=0.07,type=pdf,ext=.pdf,read=.pdf]{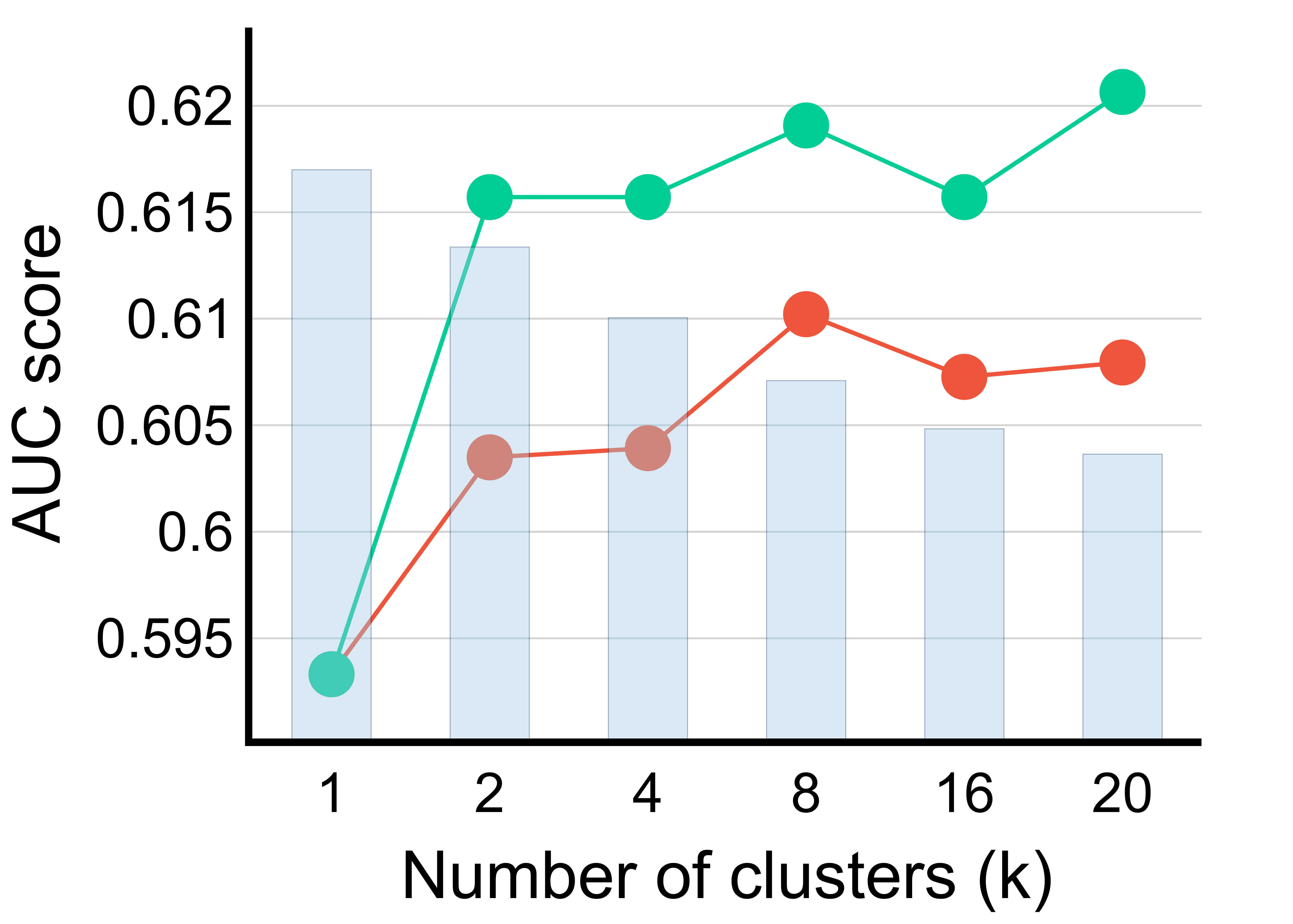} 
&
\includegraphics[scale=0.07,type=pdf,ext=.pdf,read=.pdf]{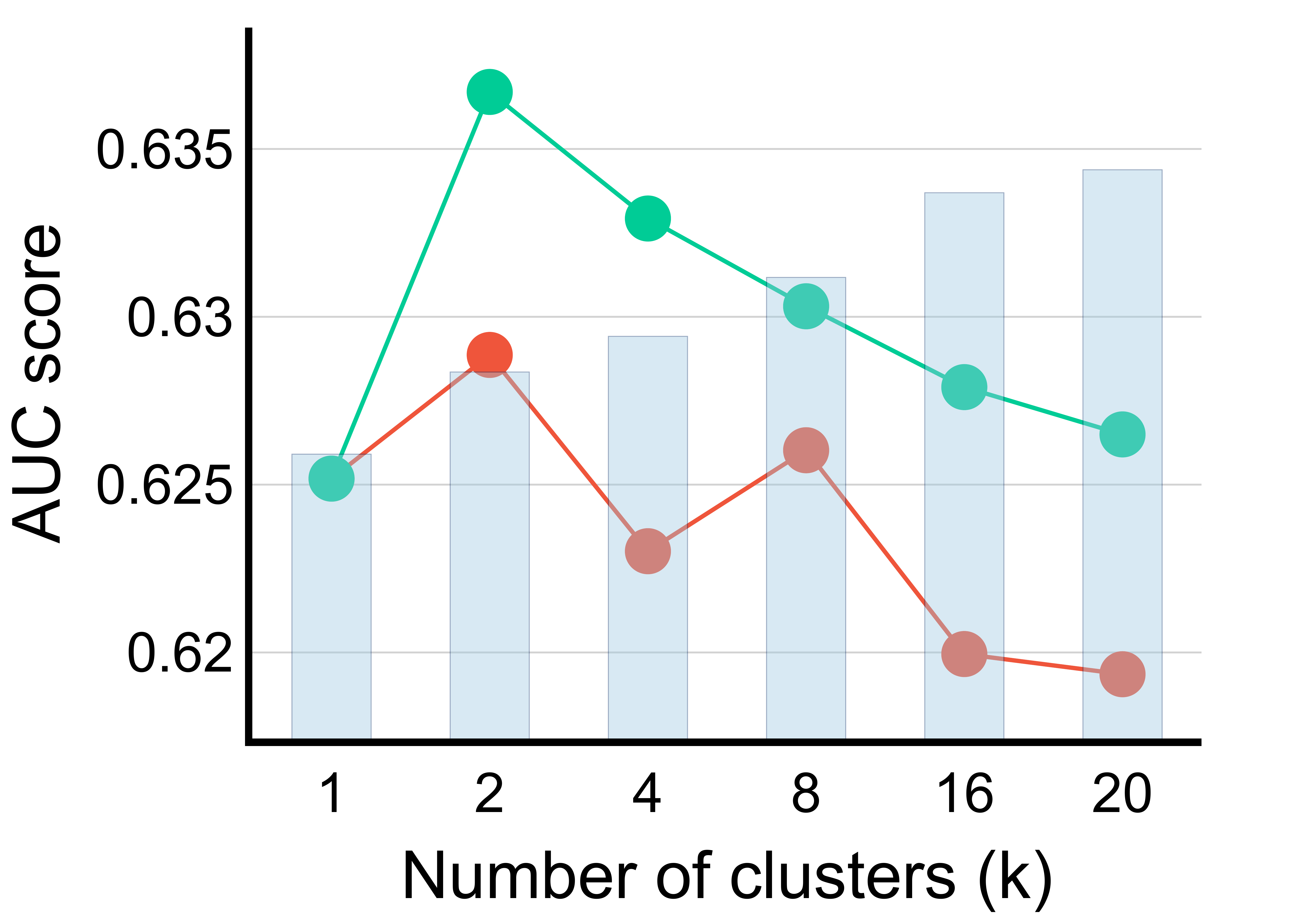} 
 & \includegraphics[scale=0.07,type=pdf,ext=.pdf,read=.pdf]{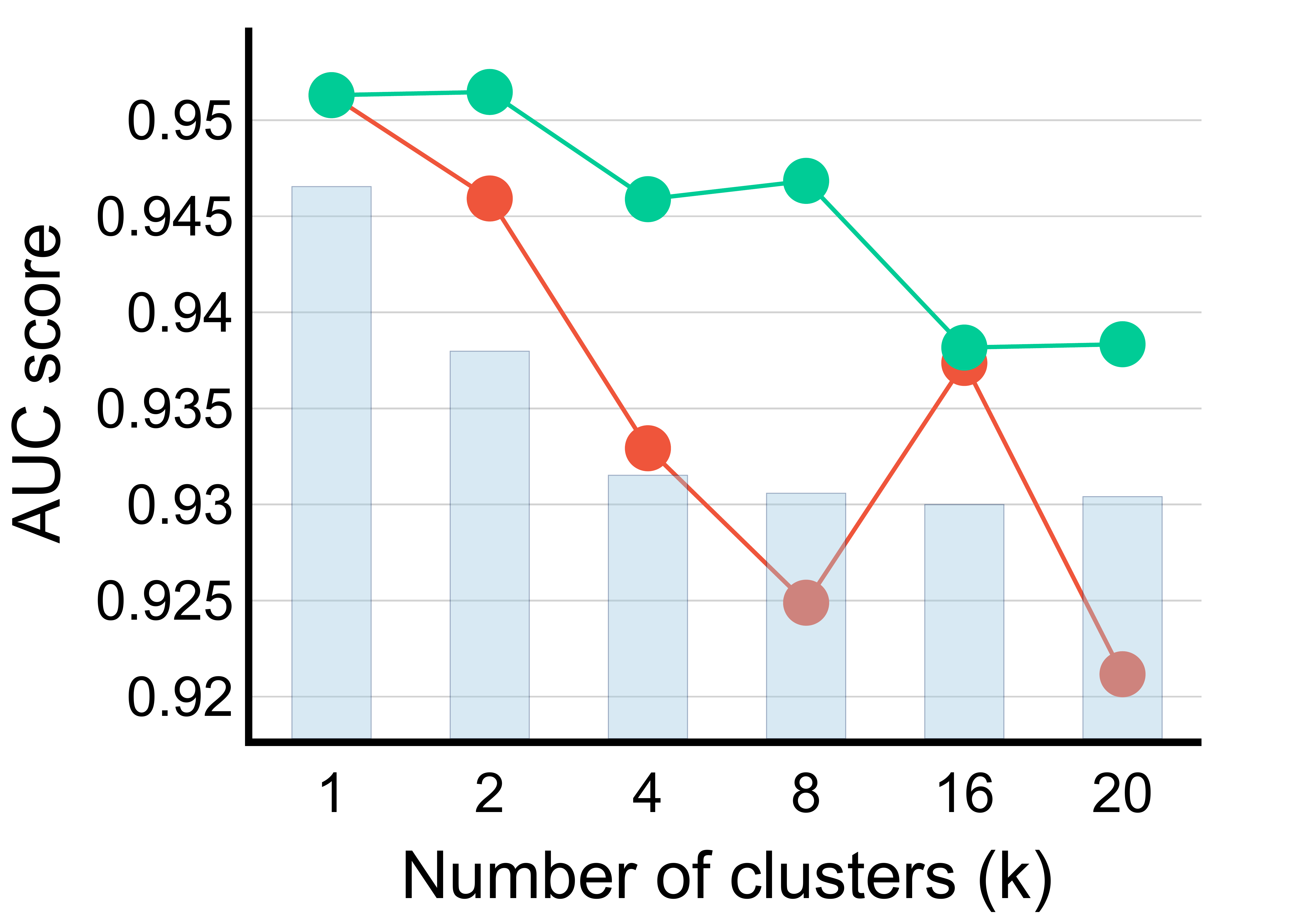}
 & \includegraphics[scale=0.07,type=pdf,ext=.pdf,read=.pdf]{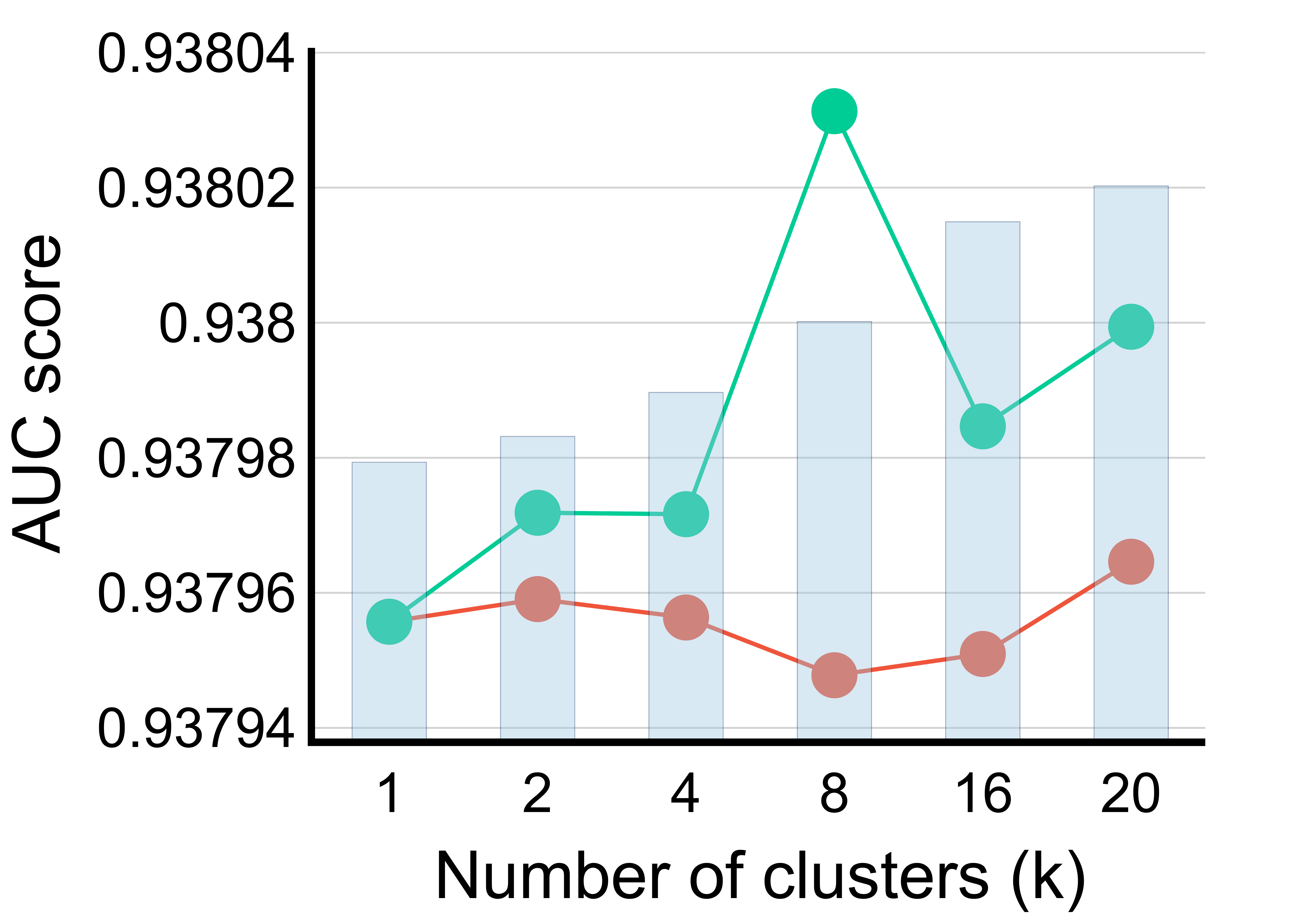}
 \includegraphics[scale=0.1]{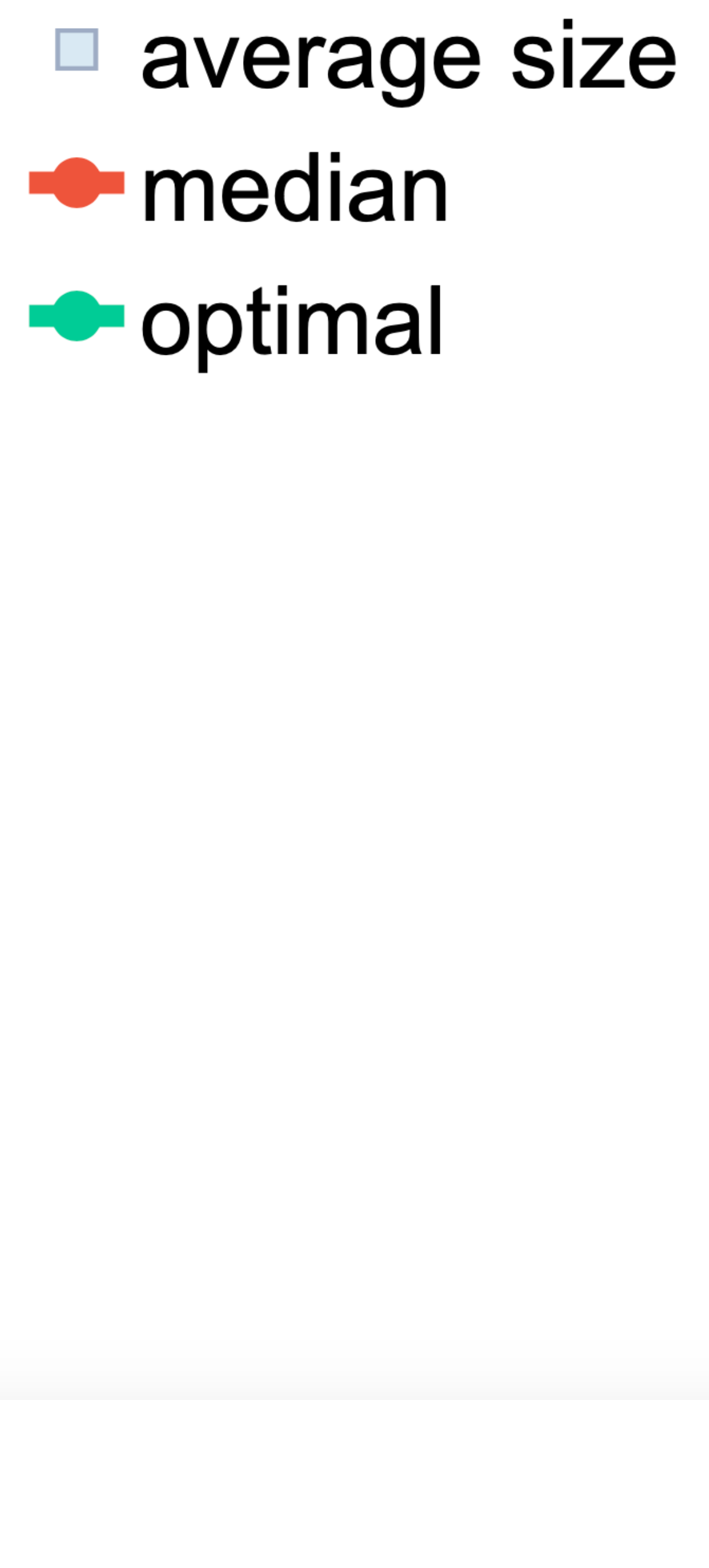}  
\\ 
\includegraphics[scale=0.07,type=pdf,ext=.pdf,read=.pdf]{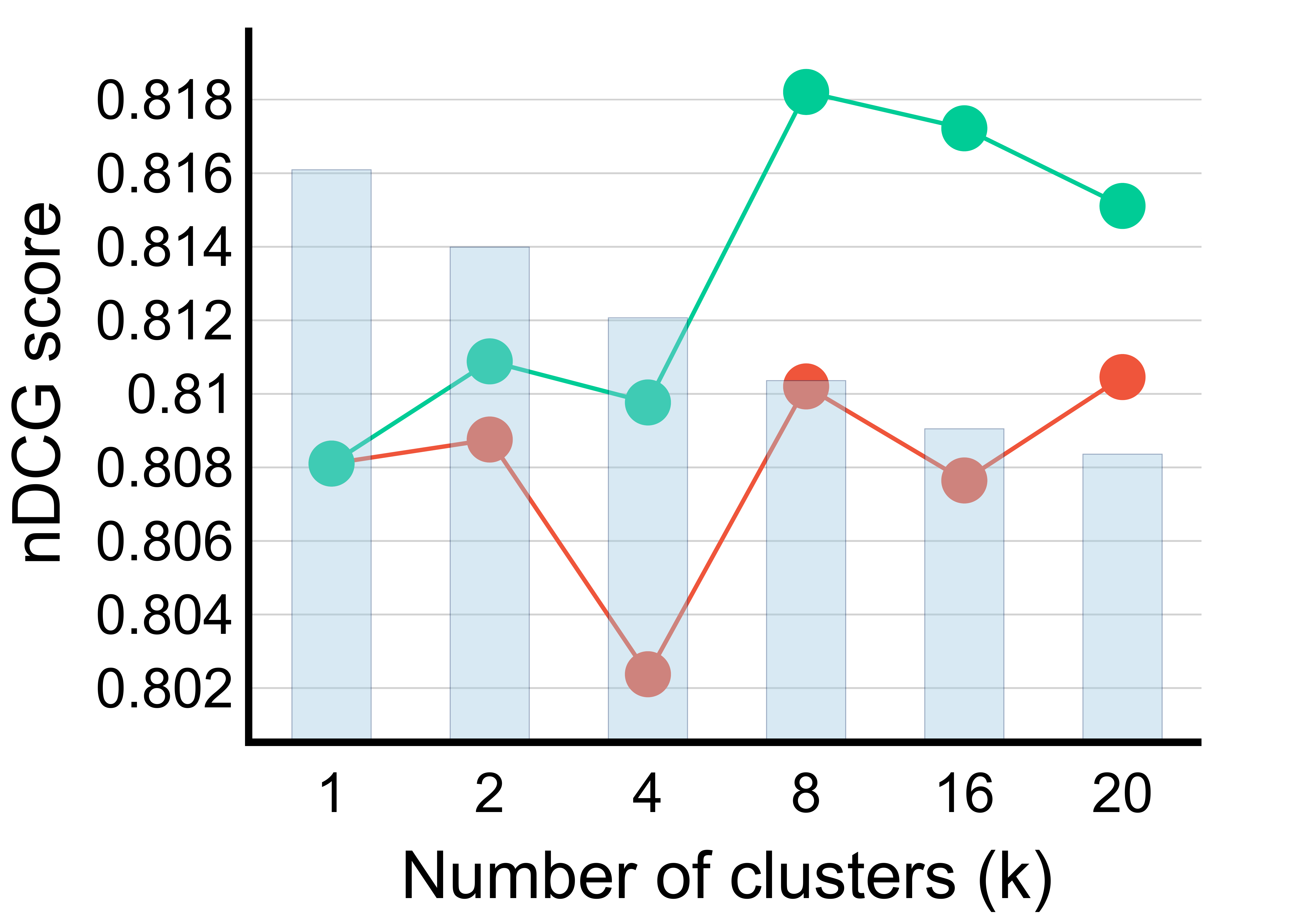} 
&
\includegraphics[scale=0.07,type=pdf,ext=.pdf,read=.pdf]{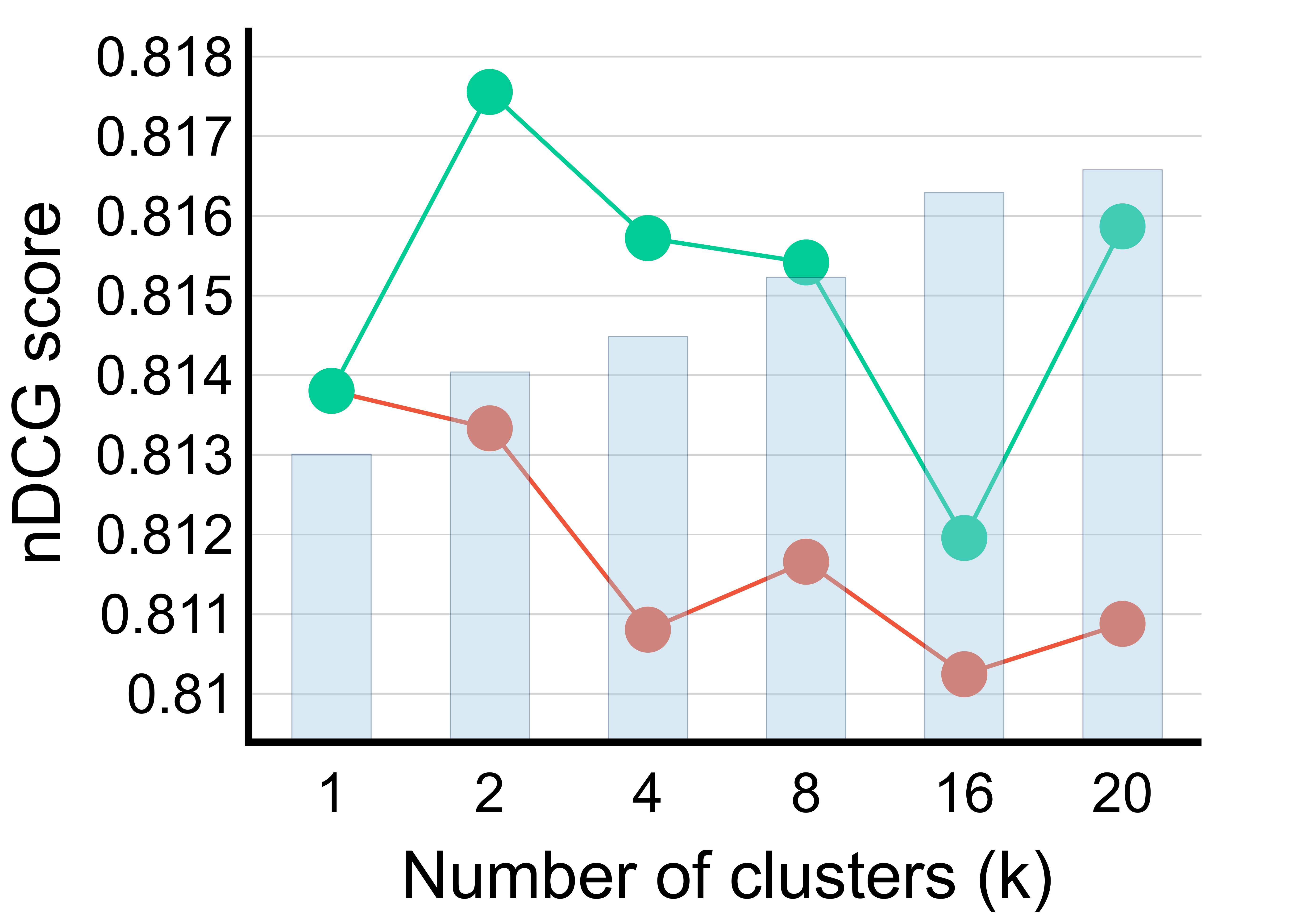} 
 & \includegraphics[scale=0.07,type=pdf,ext=.pdf,read=.pdf]{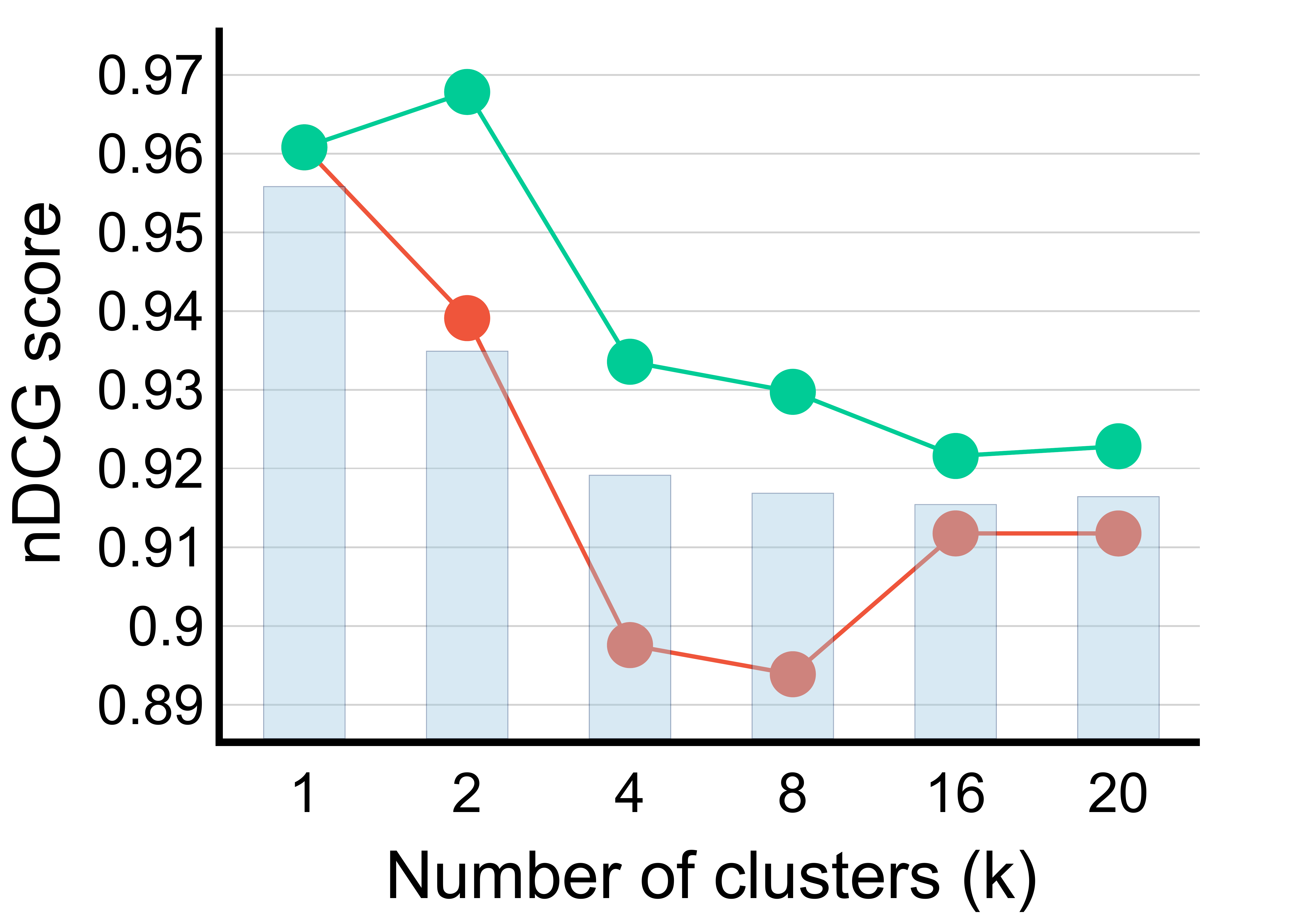}
 & \includegraphics[scale=0.07,type=pdf,ext=.pdf,read=.pdf]{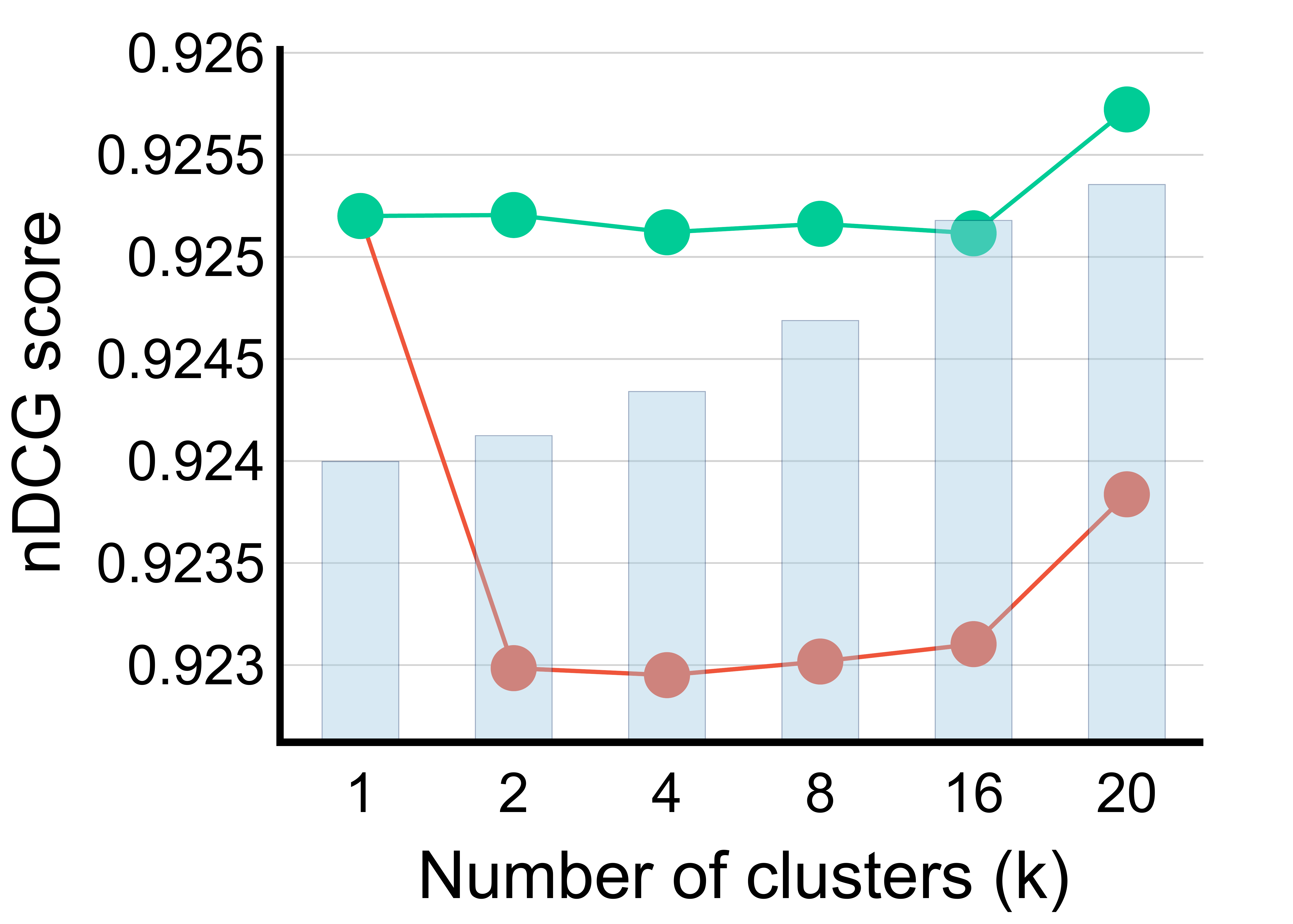}
 \includegraphics[scale=0.1]{fig/legend_varying-k.pdf}  
\end{tabular}
\caption{Compressed size and AUC or nDCG score vs. $K$ for
  Bank and Probe datasets}\label{fig:bank-probe}
\end{figure*}

\begin{figure*}[tb]
\begin{tabular}{cccc}
Windows (\AVCstar) & BSD (\AVCstar) &
 Linux (\AVCstar) & Android (\AVCstar)\\
\includegraphics[scale=0.07,type=pdf,ext=.pdf,read=.pdf]{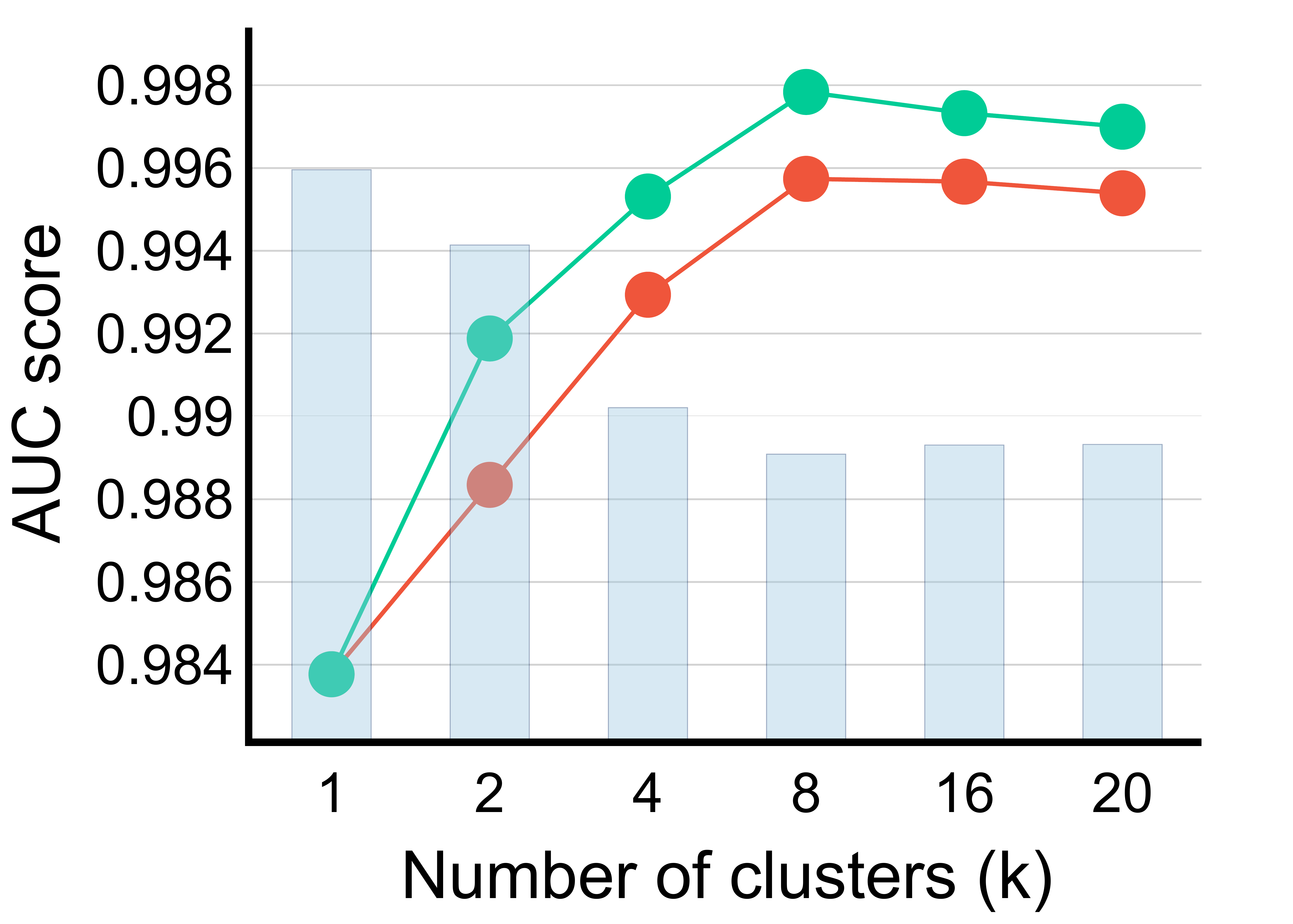} 
 & \includegraphics[scale=0.07,type=pdf,ext=.pdf,read=.pdf]{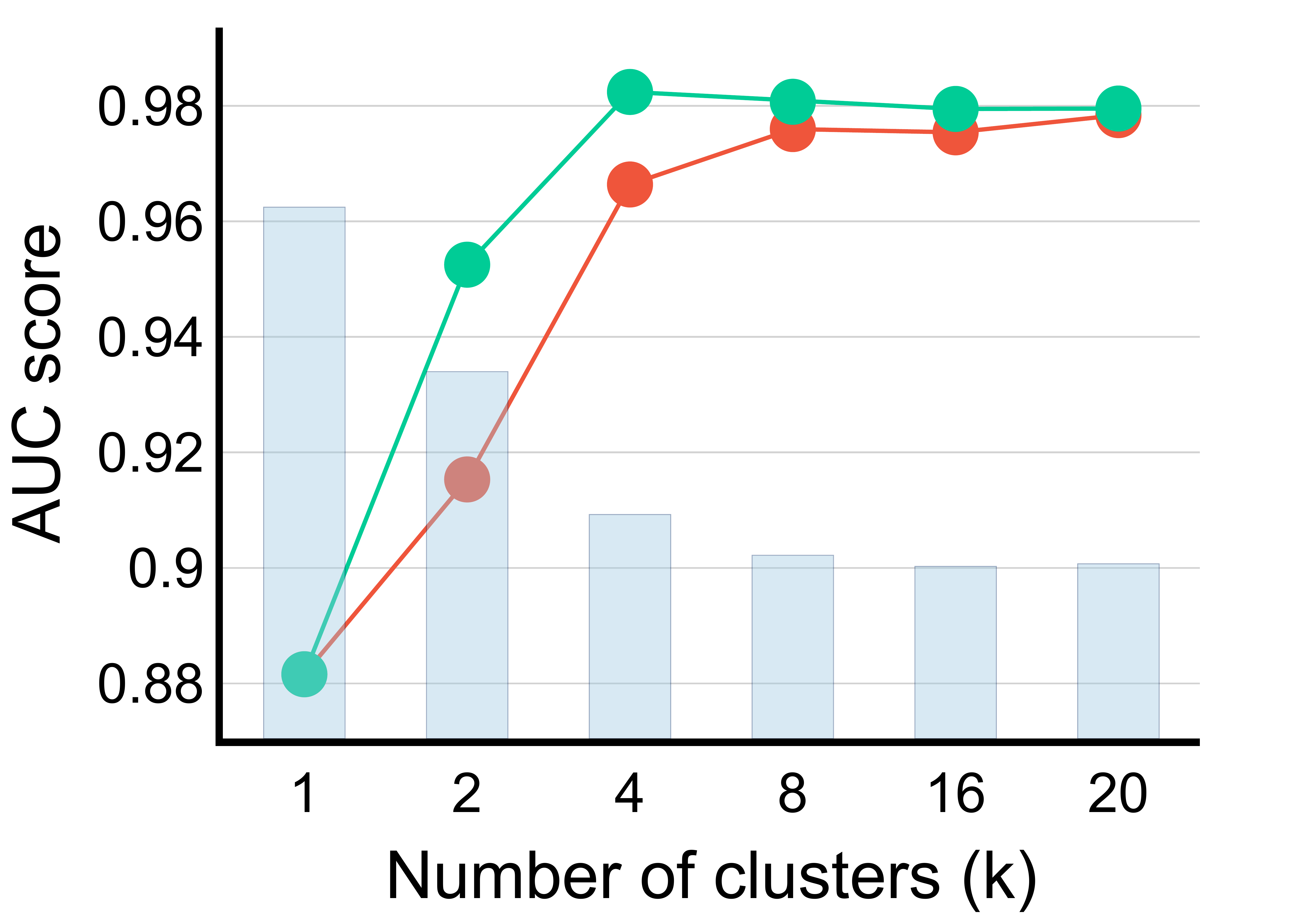} 

  & \includegraphics[scale=0.07,type=pdf,ext=.pdf,read=.pdf]{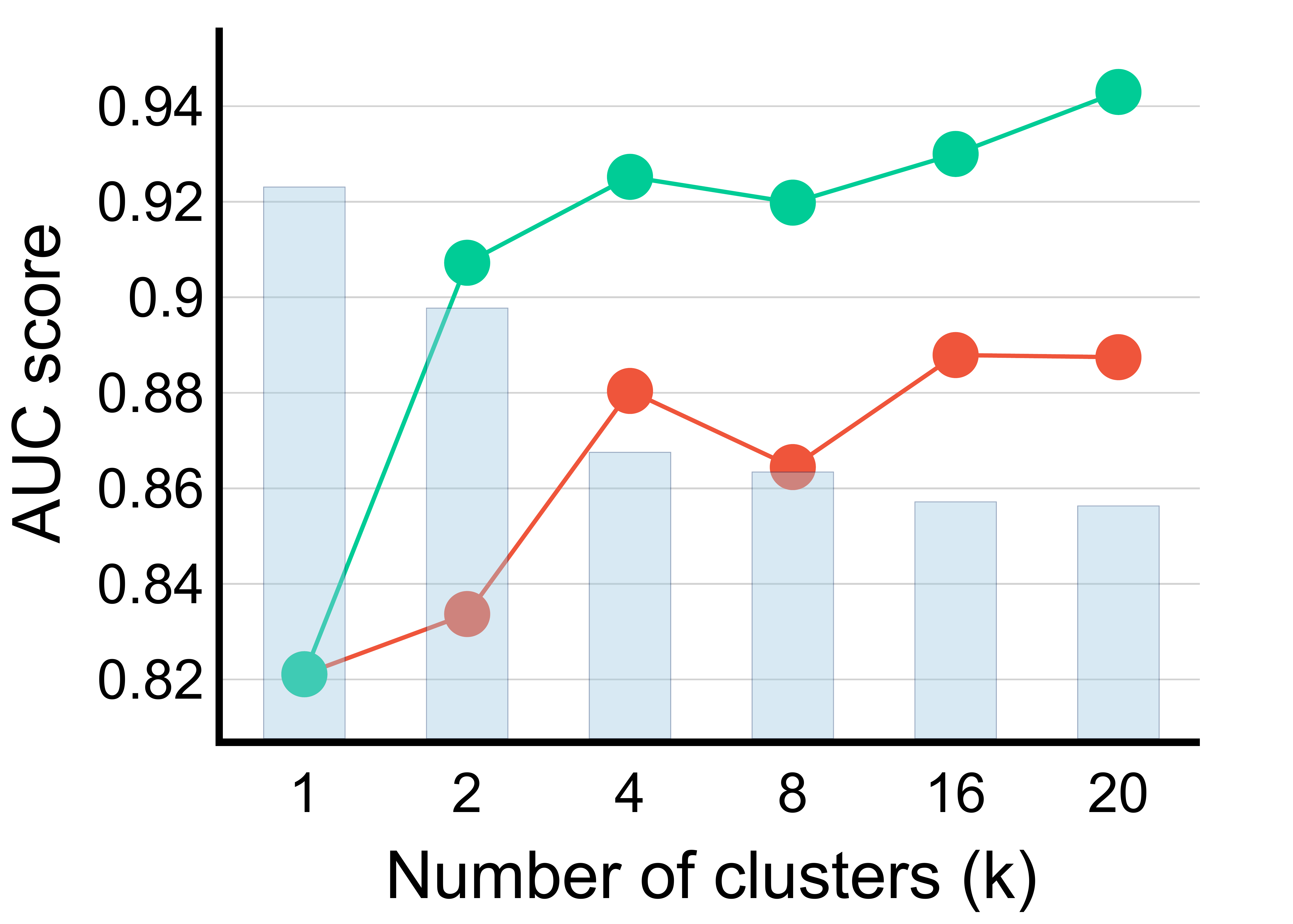} 
& \includegraphics[scale=0.07,type=pdf,ext=.pdf,read=.pdf]{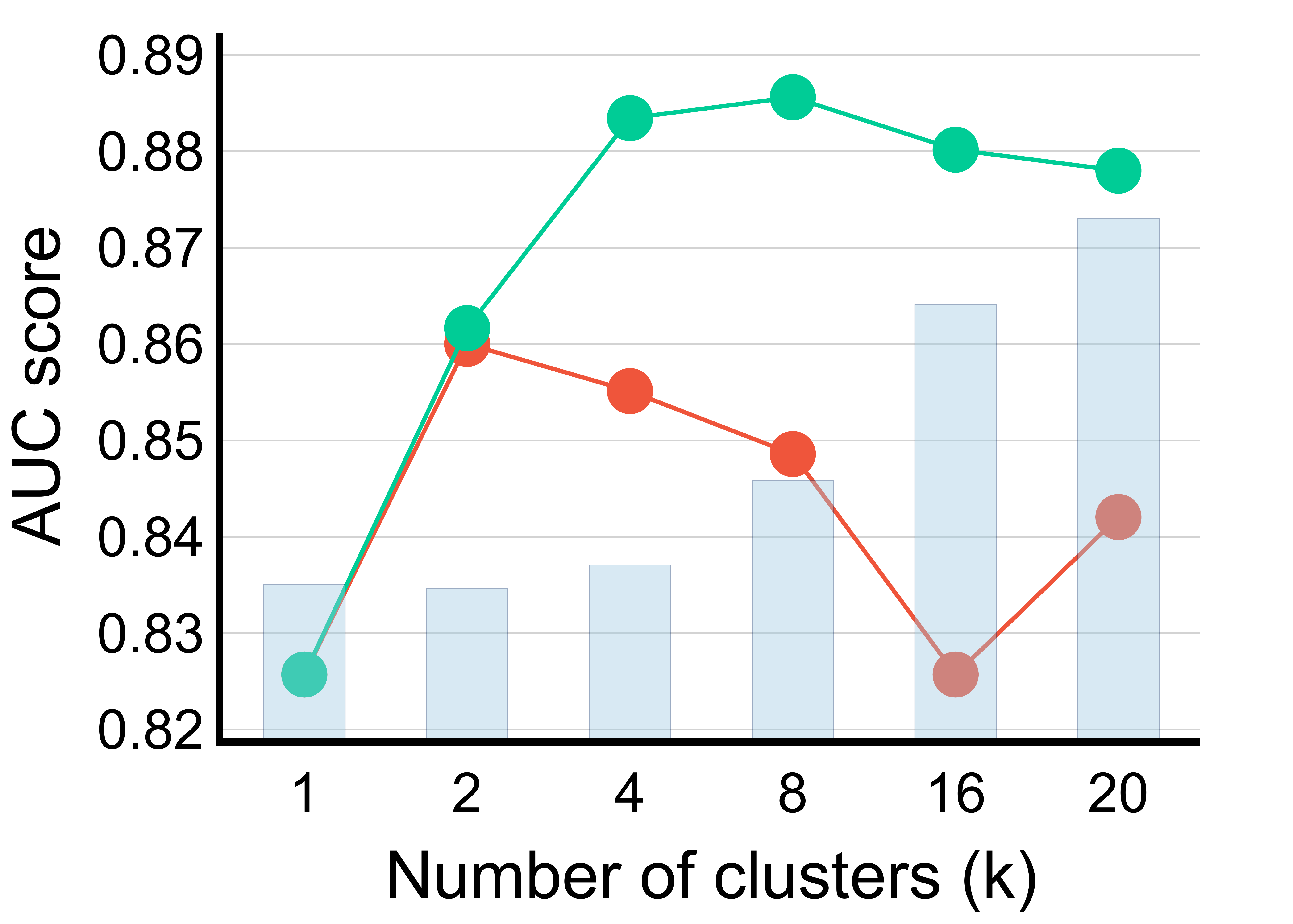} 
  \includegraphics[scale=0.1]{fig/legend_varying-k.pdf}  
\\ 
\includegraphics[scale=0.07,type=pdf,ext=.pdf,read=.pdf]{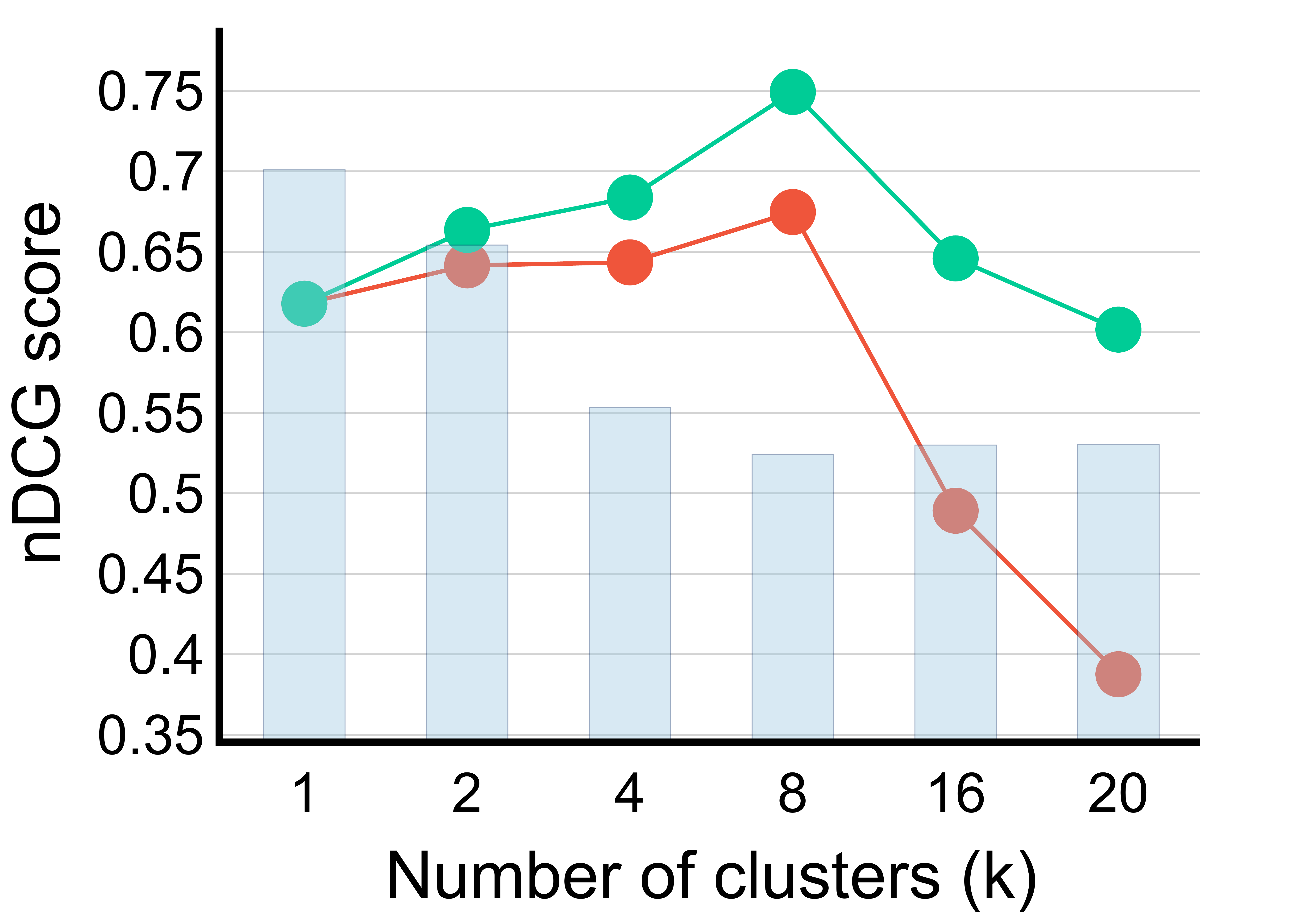} 
 & \includegraphics[scale=0.07,type=pdf,ext=.pdf,read=.pdf]{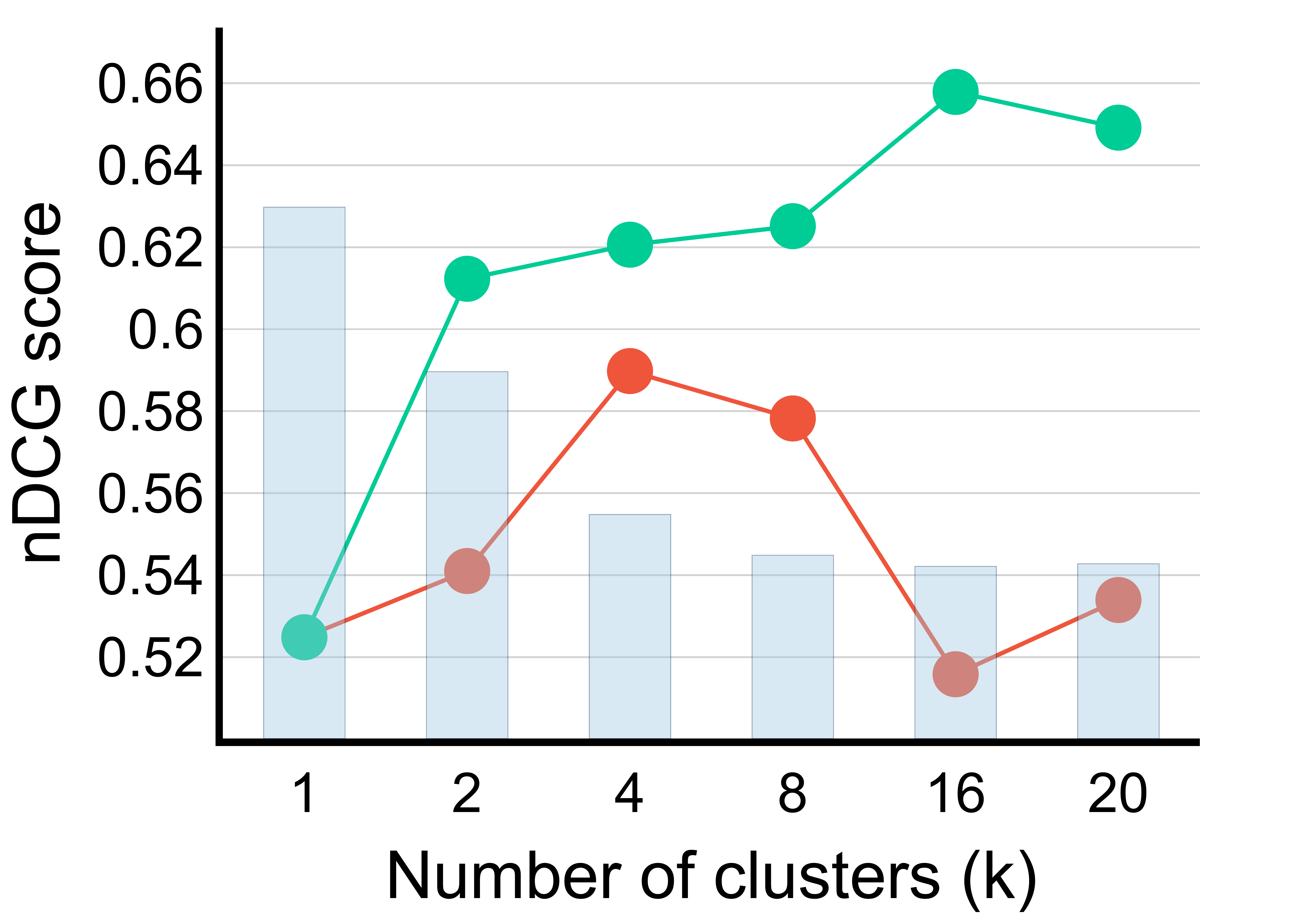} 

  & \includegraphics[scale=0.07,type=pdf,ext=.pdf,read=.pdf]{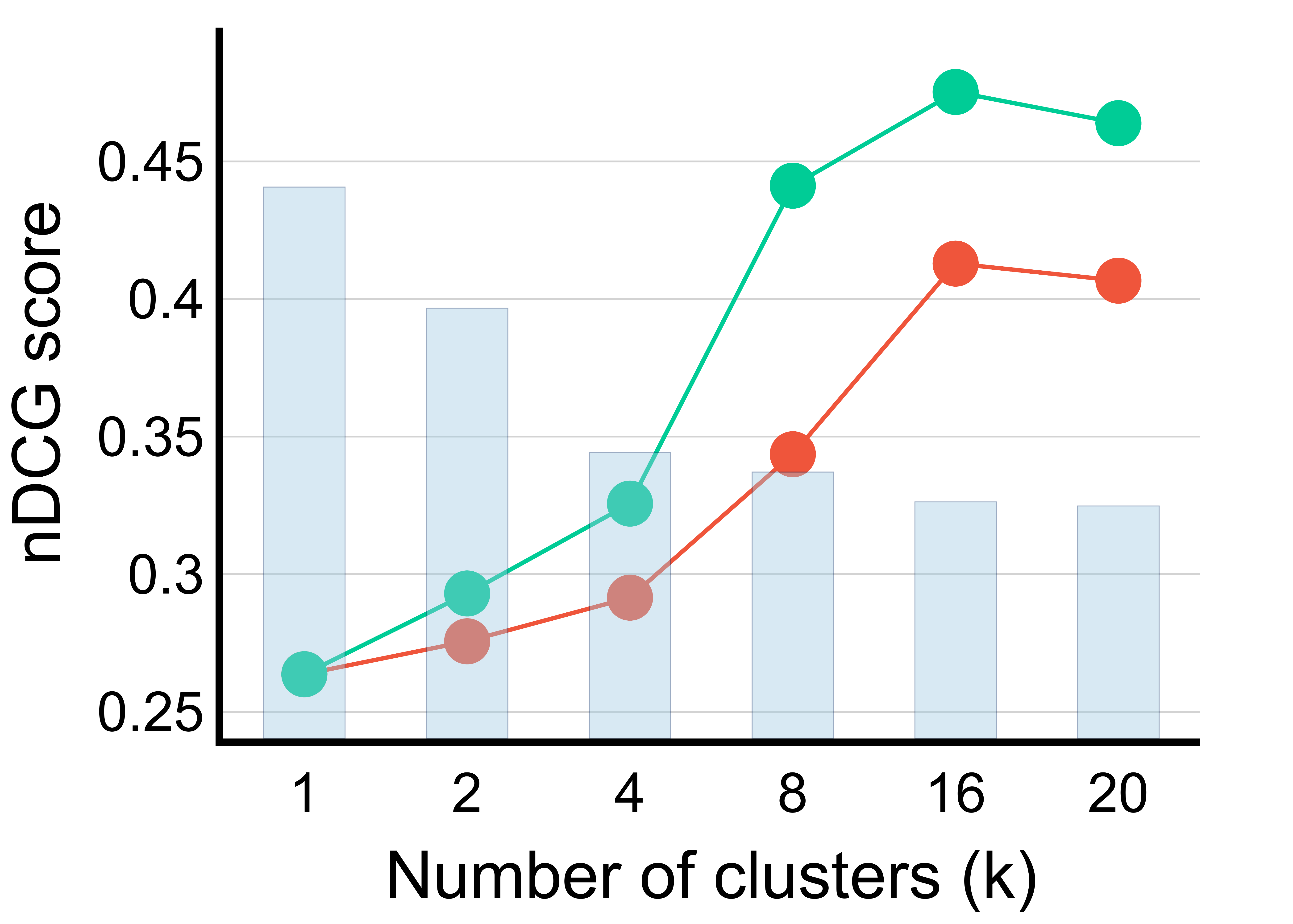} 
& \includegraphics[scale=0.07,type=pdf,ext=.pdf,read=.pdf]{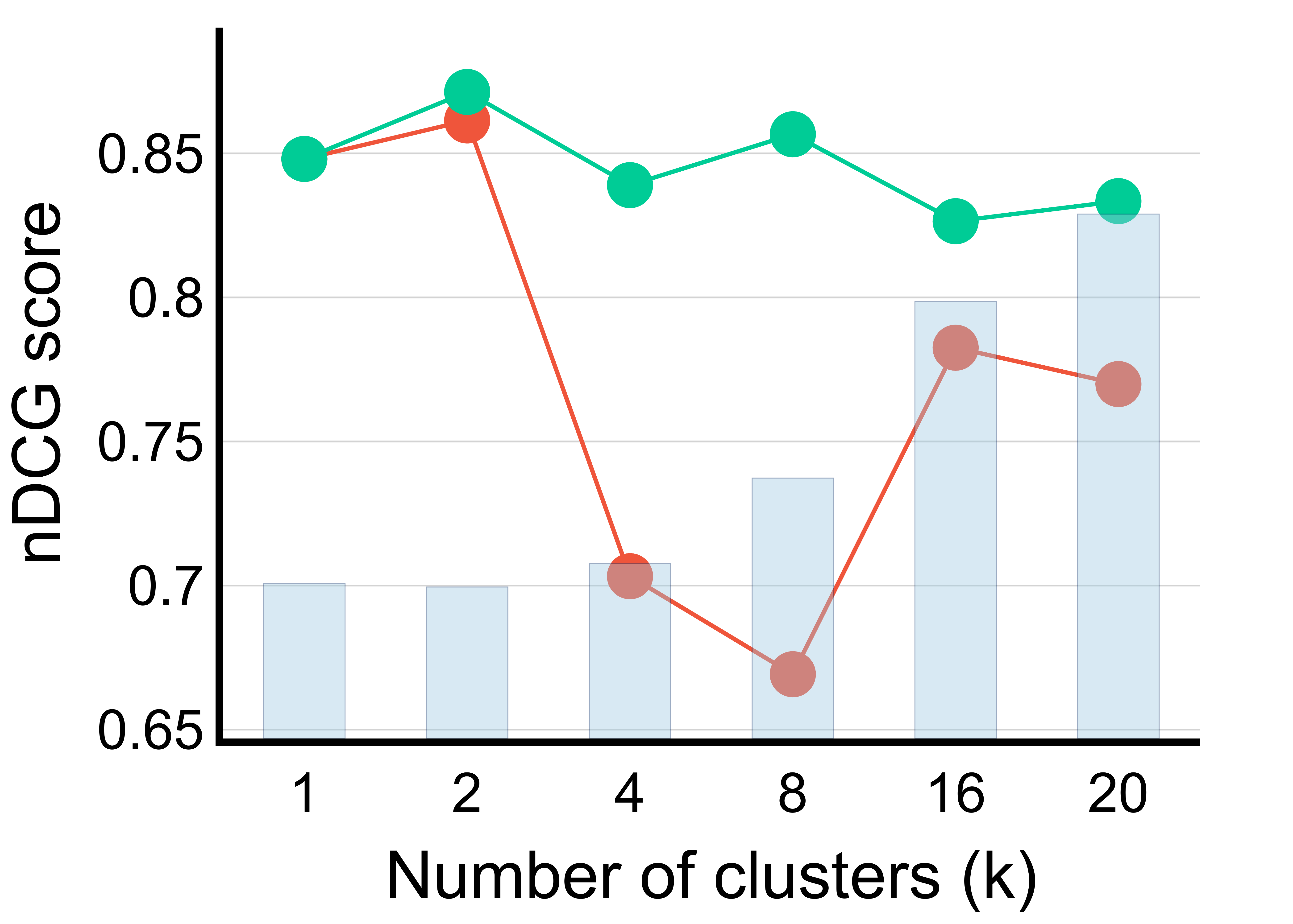} 
  \includegraphics[scale=0.1]{fig/legend_varying-k.pdf}  
\\
Windows (\Krimpstar) & BSD (\Krimpstar) & Linux (\Krimpstar)& Android (\Krimpstar)\\
\includegraphics[scale=0.07,type=pdf,ext=.pdf,read=.pdf]{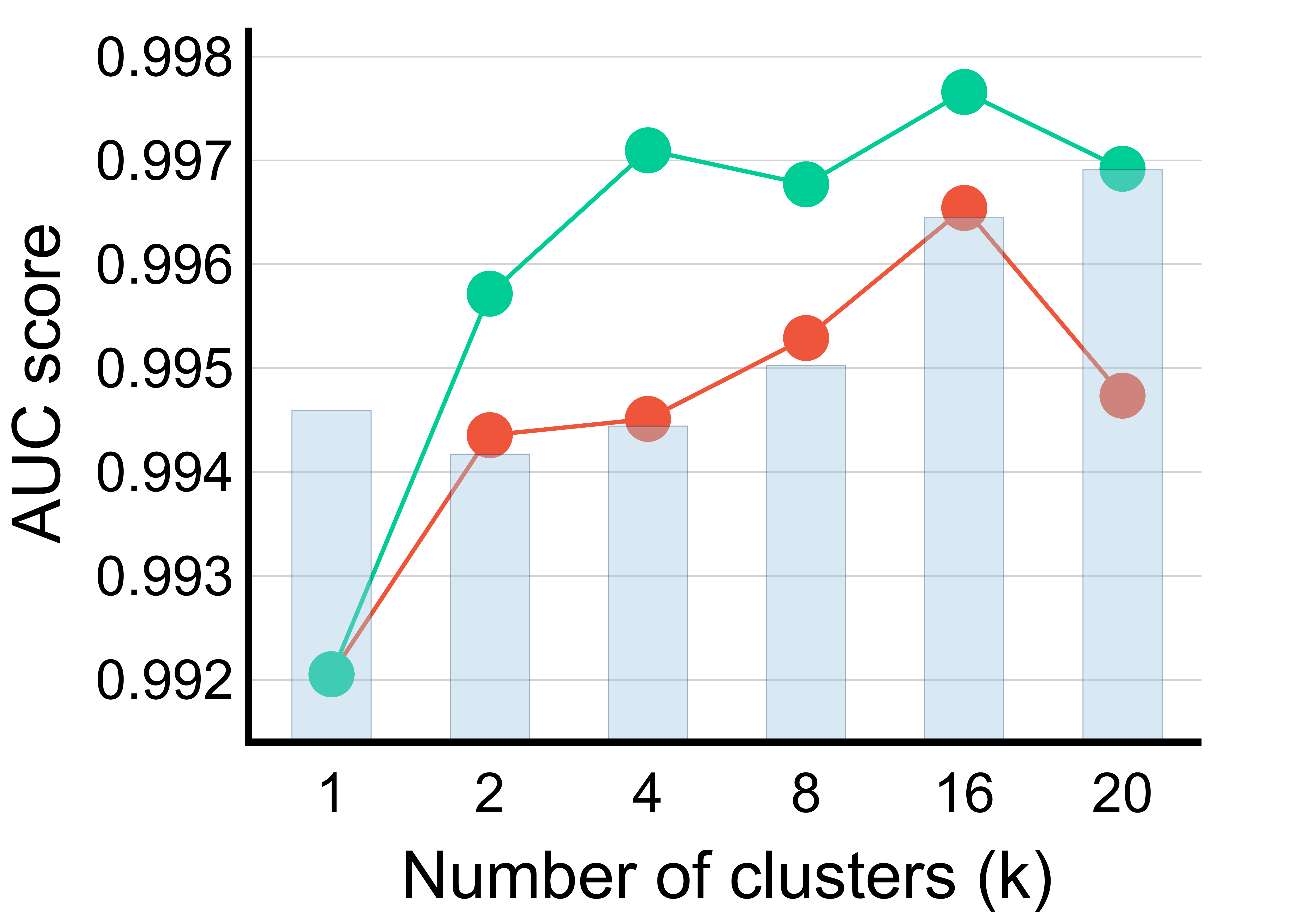} 
 & \includegraphics[scale=0.07,type=pdf,ext=.pdf,read=.pdf]{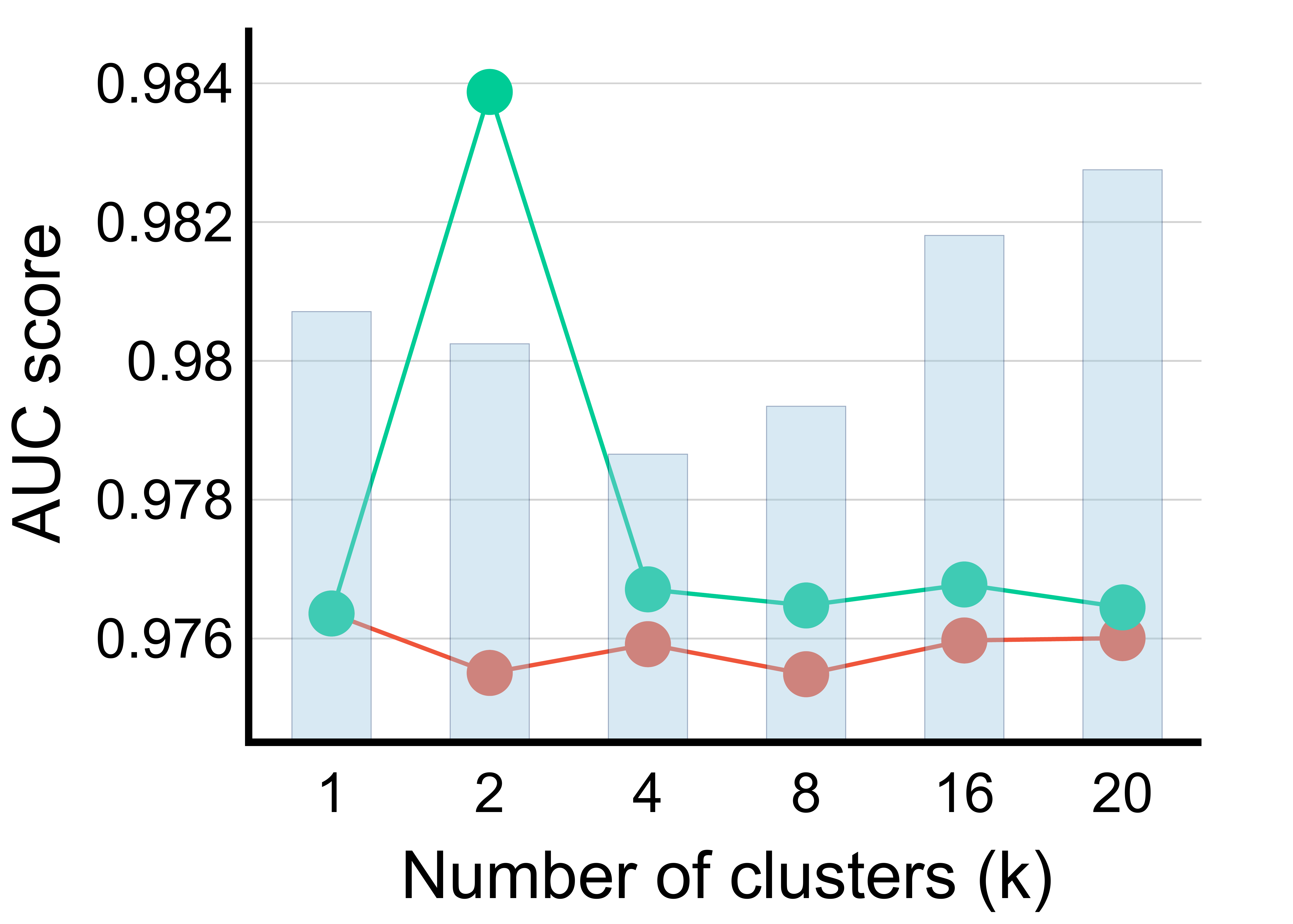} 

  & \includegraphics[scale=0.07,type=pdf,ext=.pdf,read=.pdf]{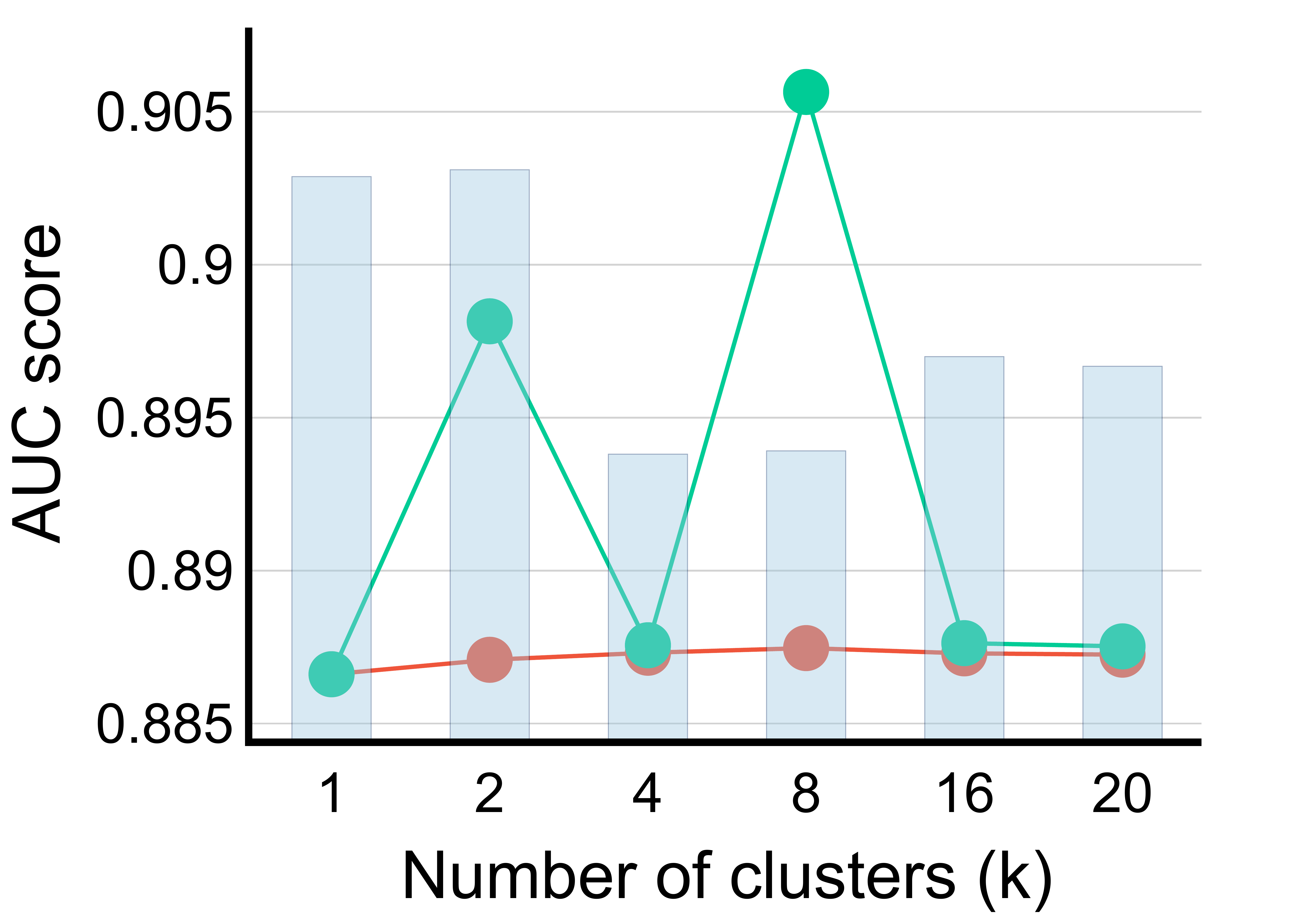} 
& \includegraphics[scale=0.07,type=pdf,ext=.pdf,read=.pdf]{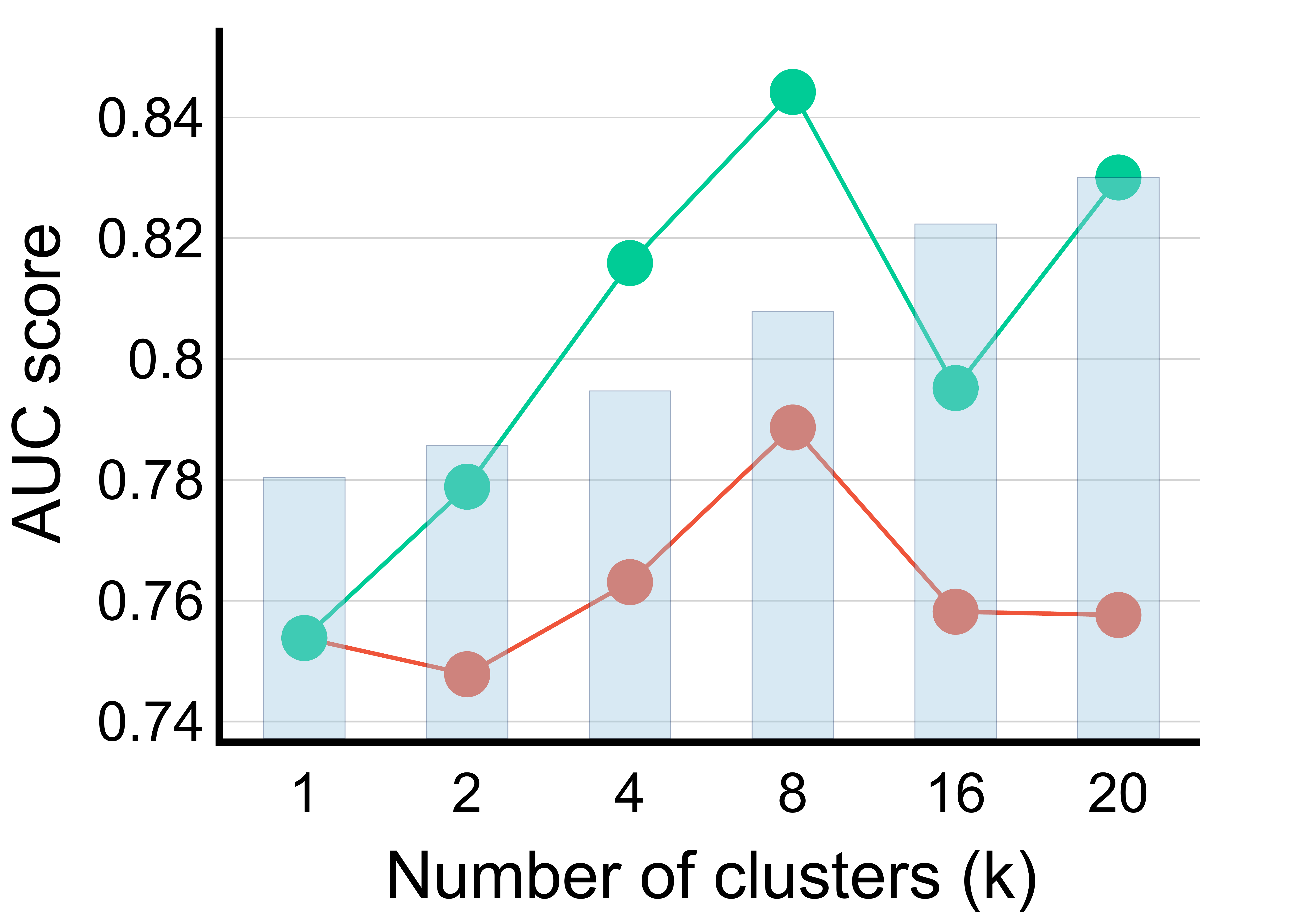} 
  \includegraphics[scale=0.1]{fig/legend_varying-k.pdf}  
\\ 
\includegraphics[scale=0.07,type=pdf,ext=.pdf,read=.pdf]{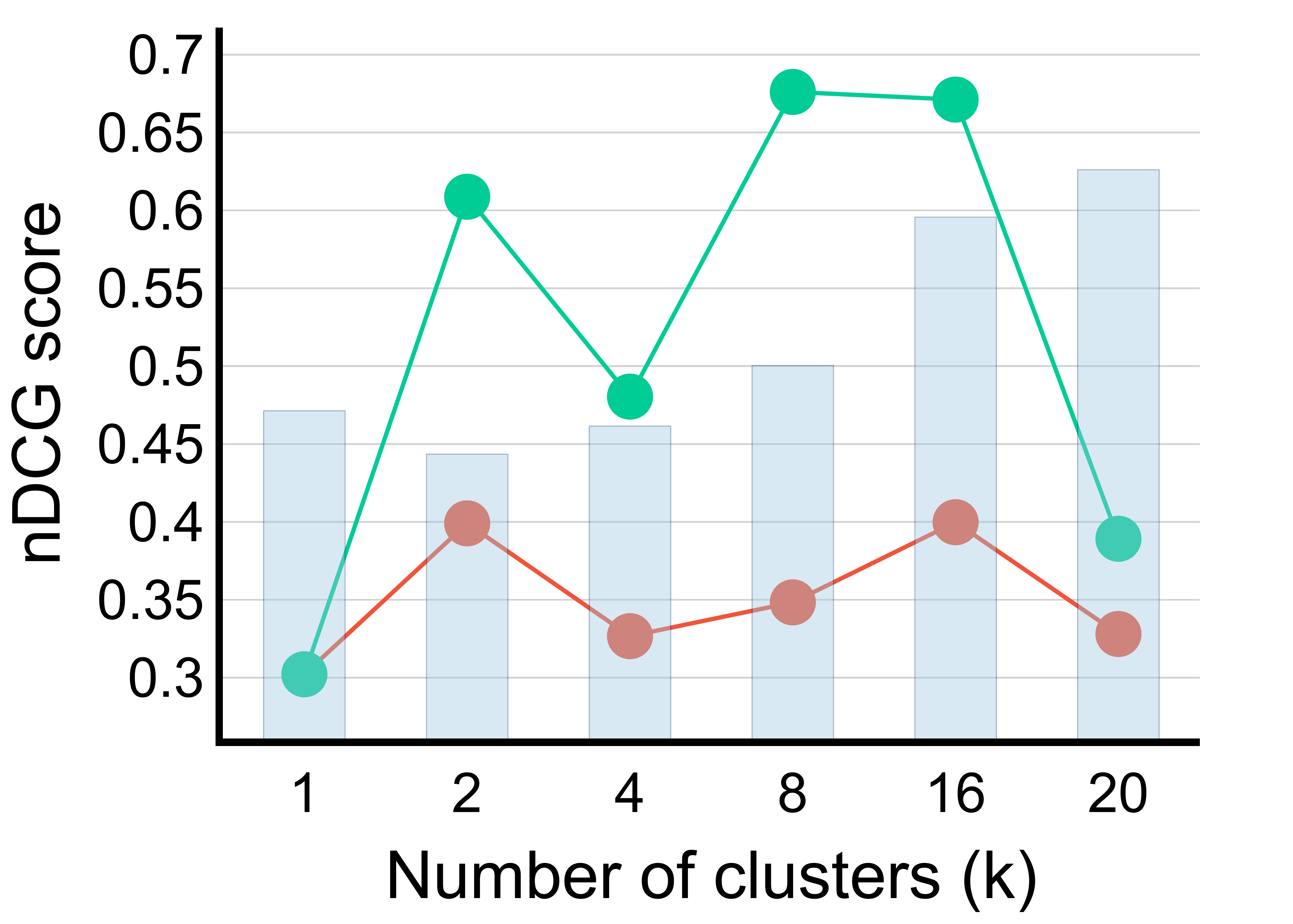} 
 & \includegraphics[scale=0.07,type=pdf,ext=.pdf,read=.pdf]{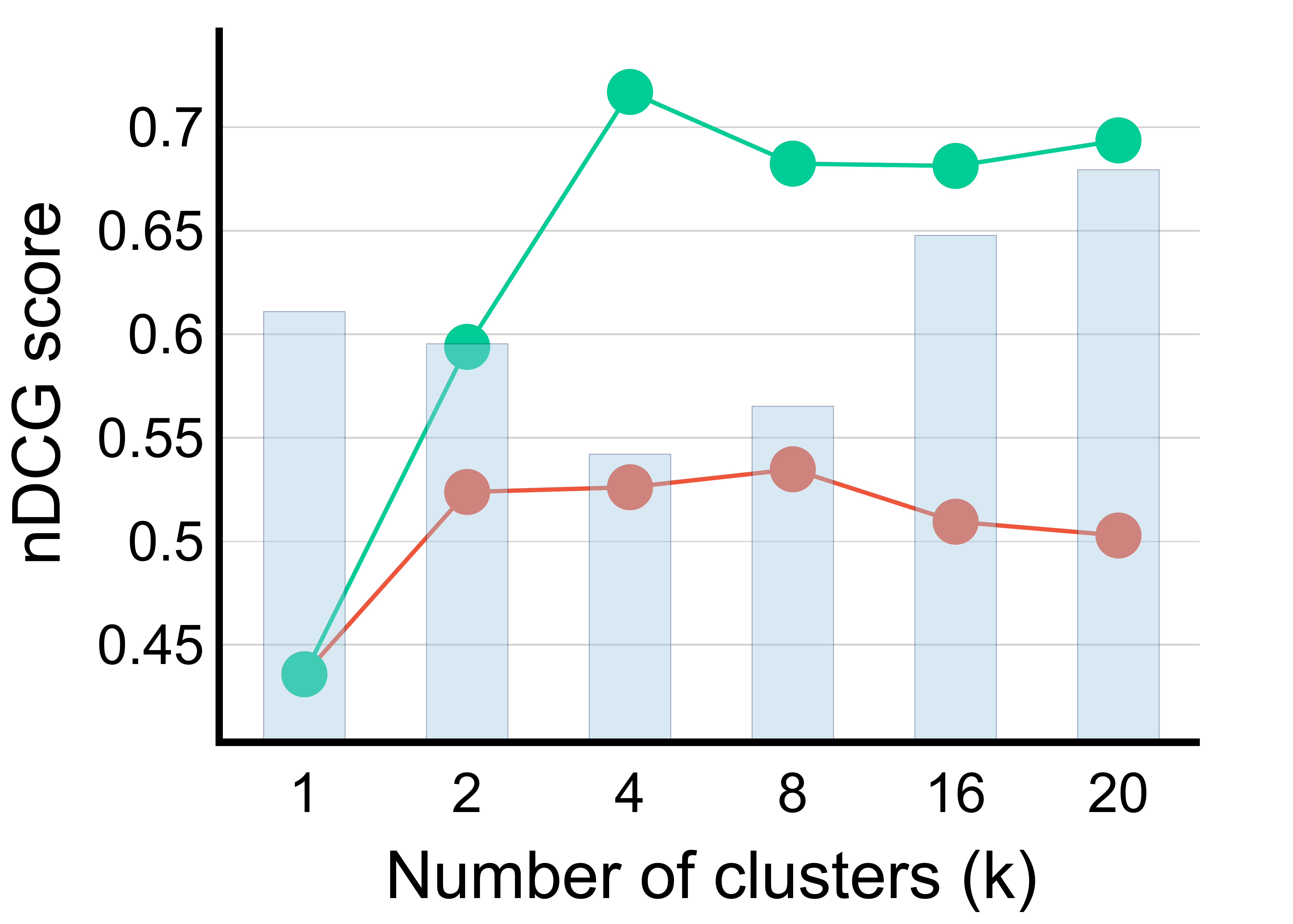} 

  & \includegraphics[scale=0.07,type=pdf,ext=.pdf,read=.pdf]{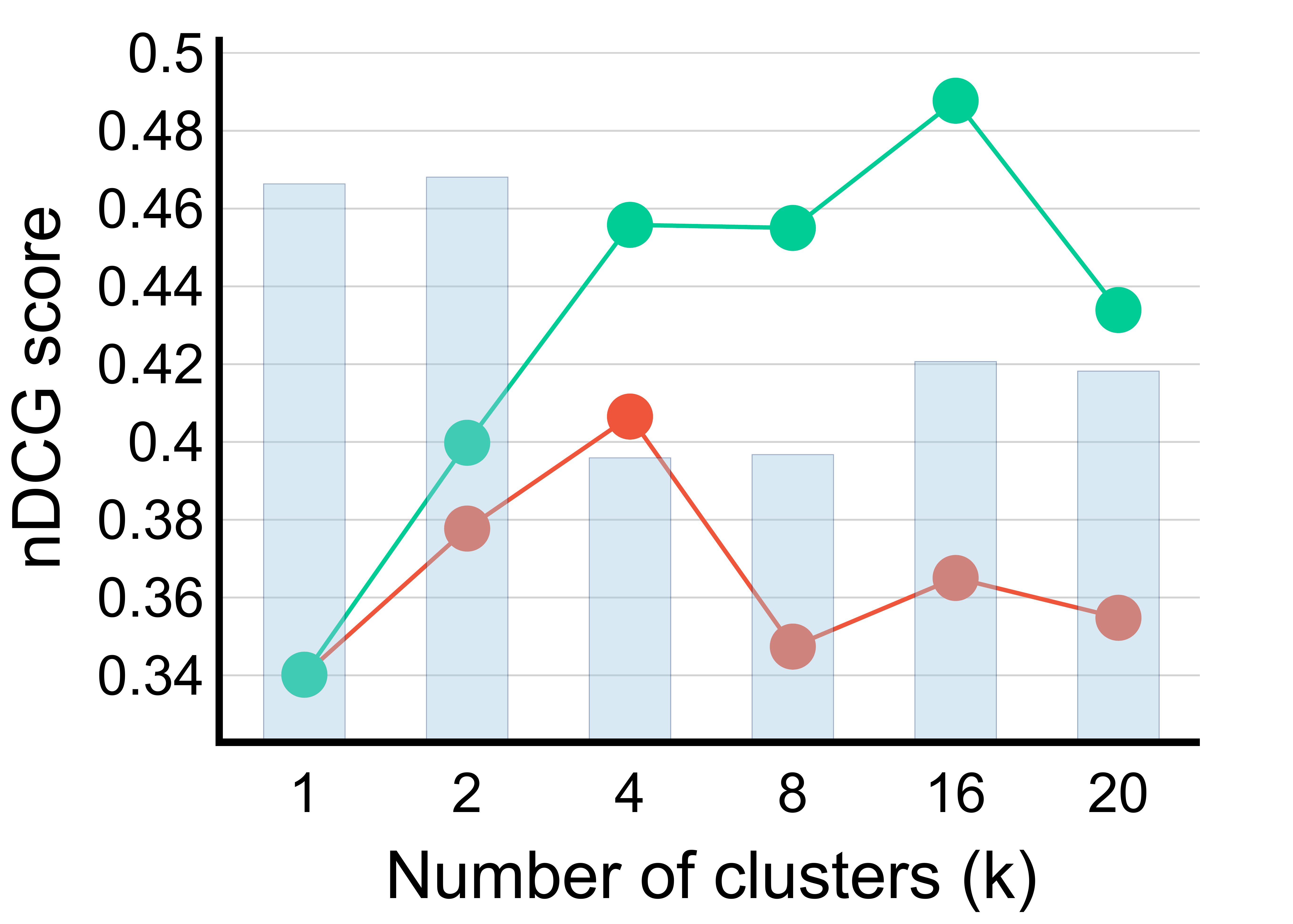} 
& \includegraphics[scale=0.07,type=pdf,ext=.pdf,read=.pdf]{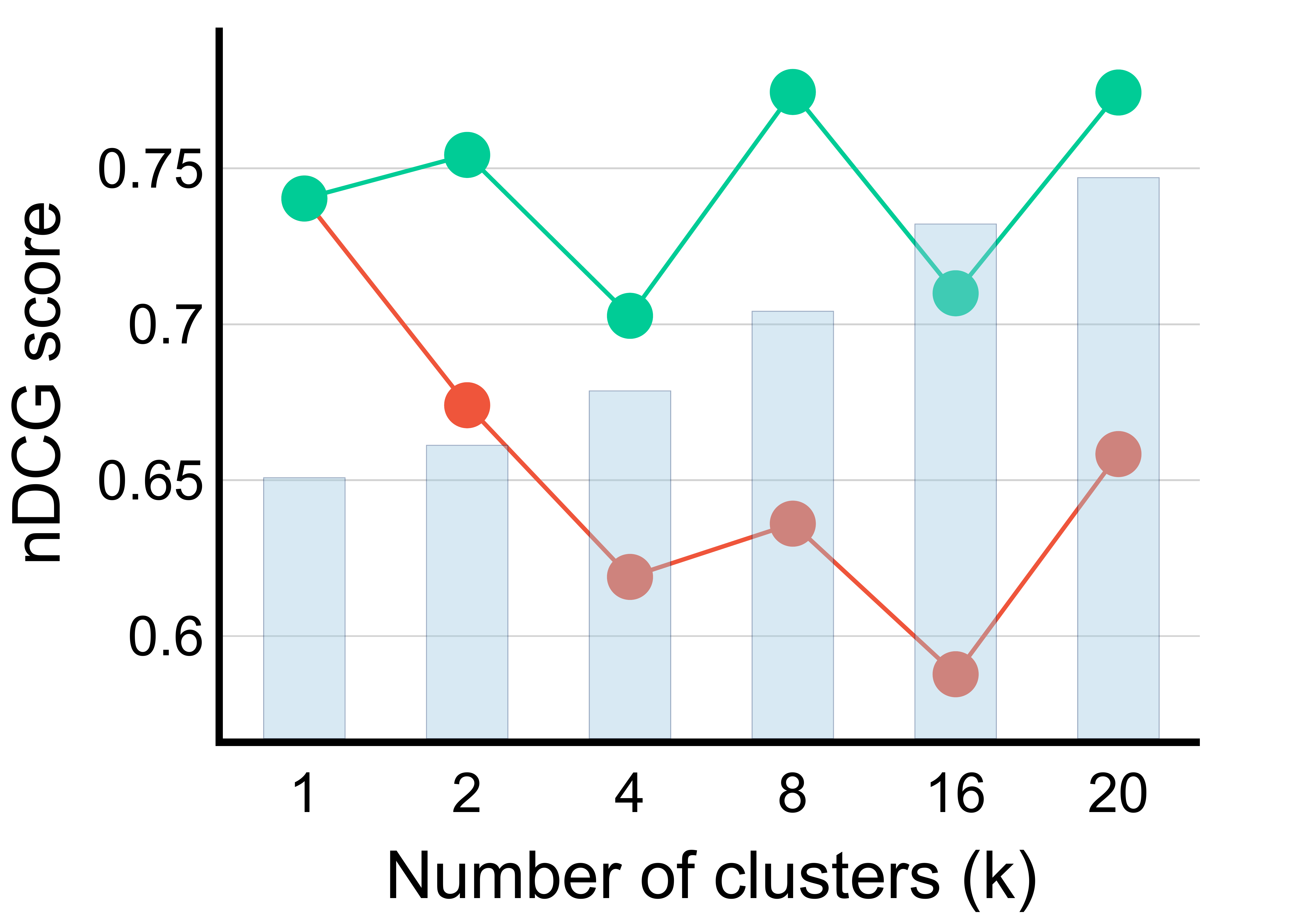} 
  \includegraphics[scale=0.1]{fig/legend_varying-k.pdf}  
\\
Windows (\CompreXstar) & BSD (\CompreXstar) & Linux (\CompreXstar)& Android (\CompreXstar)\\
\includegraphics[scale=0.07,type=pdf,ext=.pdf,read=.pdf]{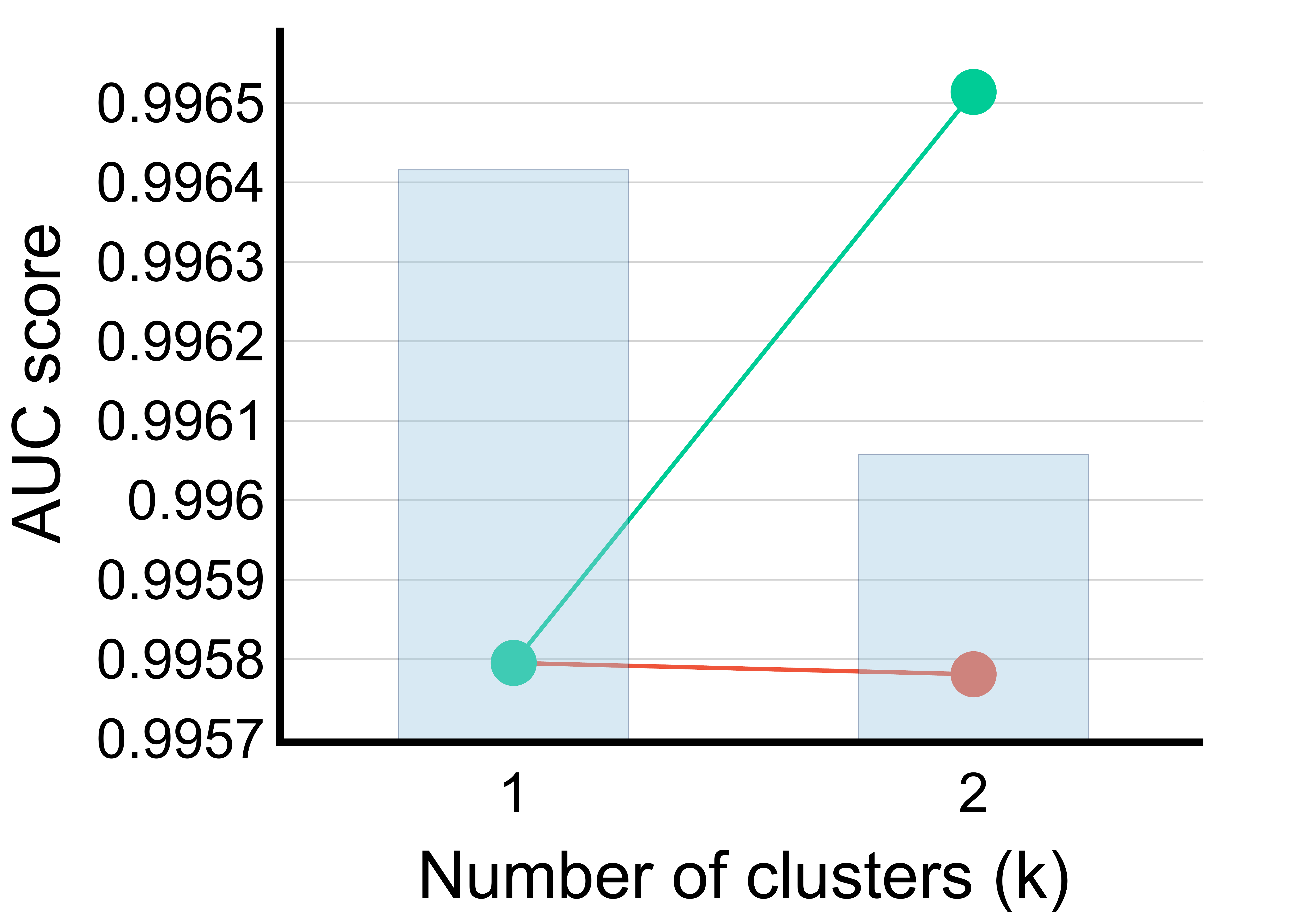} 
 & \includegraphics[scale=0.07,type=pdf,ext=.pdf,read=.pdf]{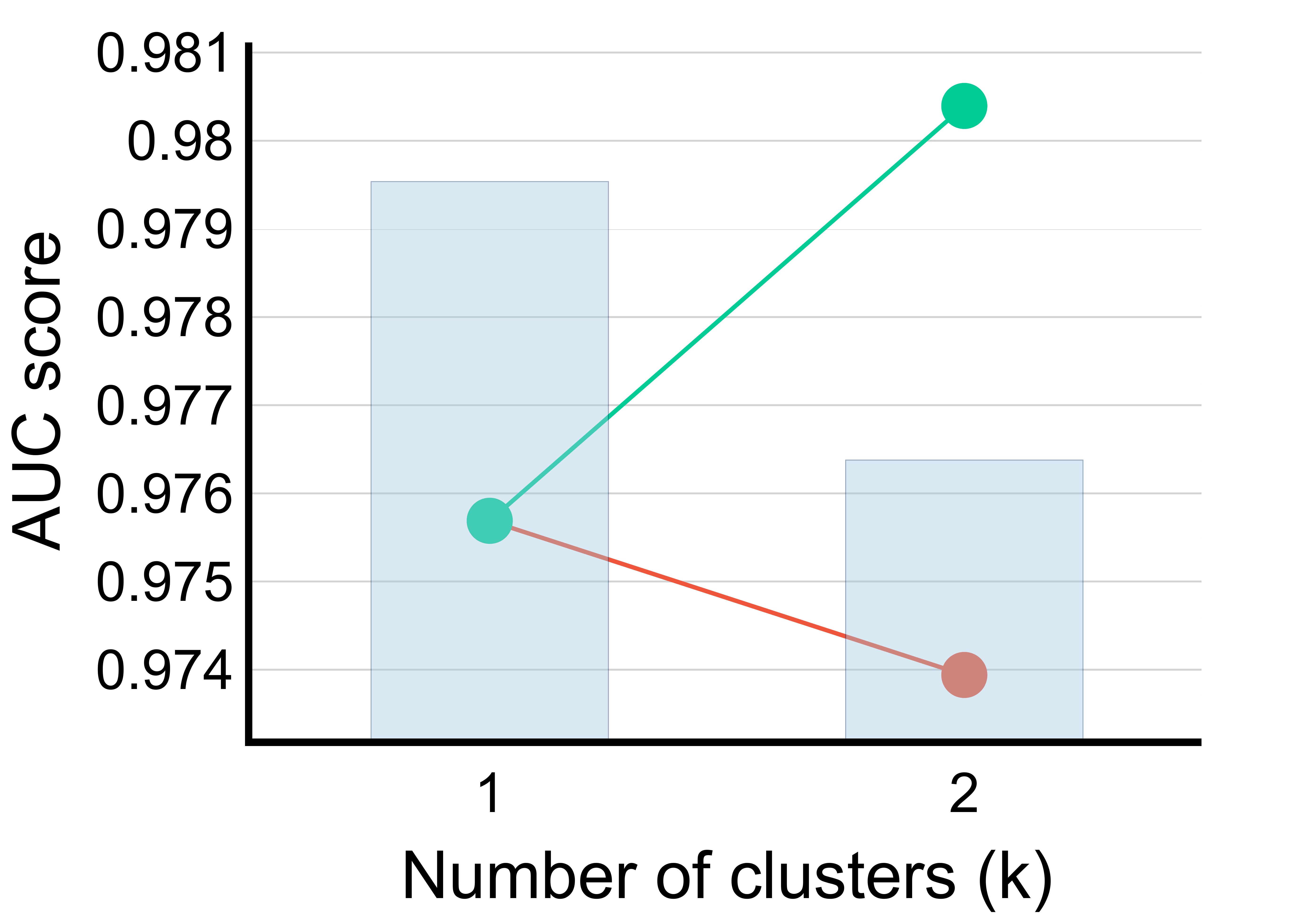} 

  & \includegraphics[scale=0.07,type=pdf,ext=.pdf,read=.pdf]{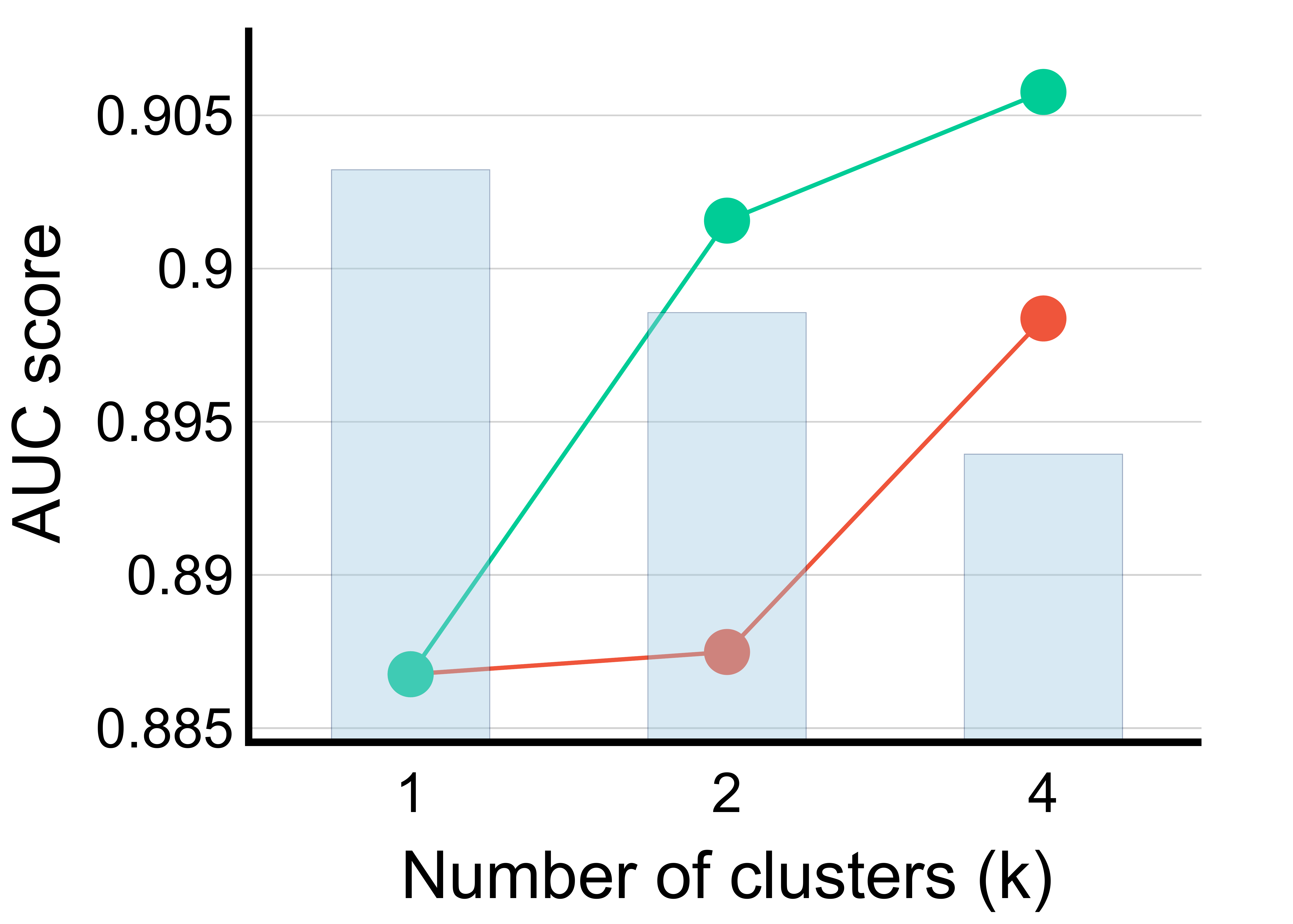} 
& \includegraphics[scale=0.07,type=pdf,ext=.pdf,read=.pdf]{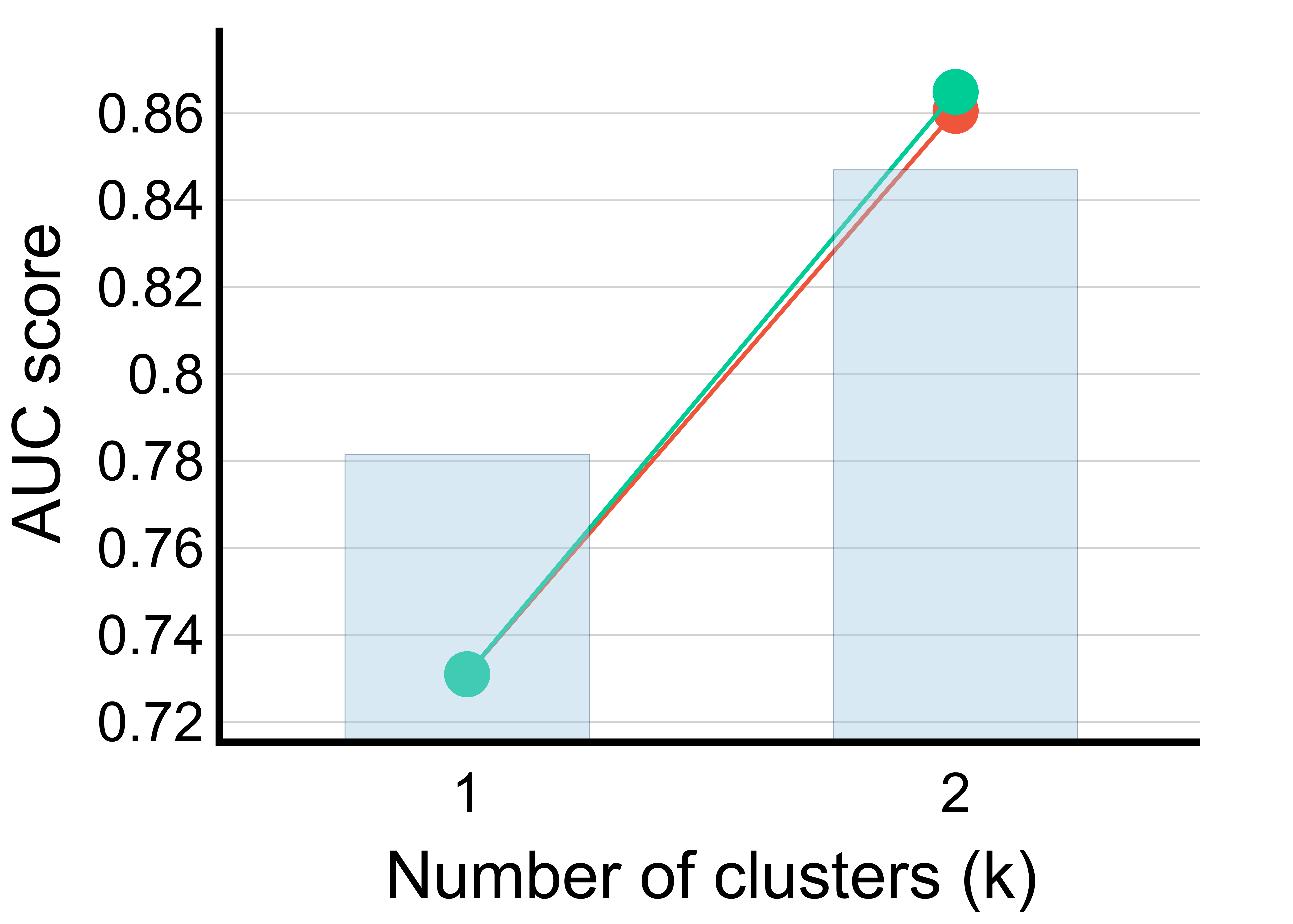} 
  \includegraphics[scale=0.1]{fig/legend_varying-k.pdf}  
\\ 
\includegraphics[scale=0.07,type=pdf,ext=.pdf,read=.pdf]{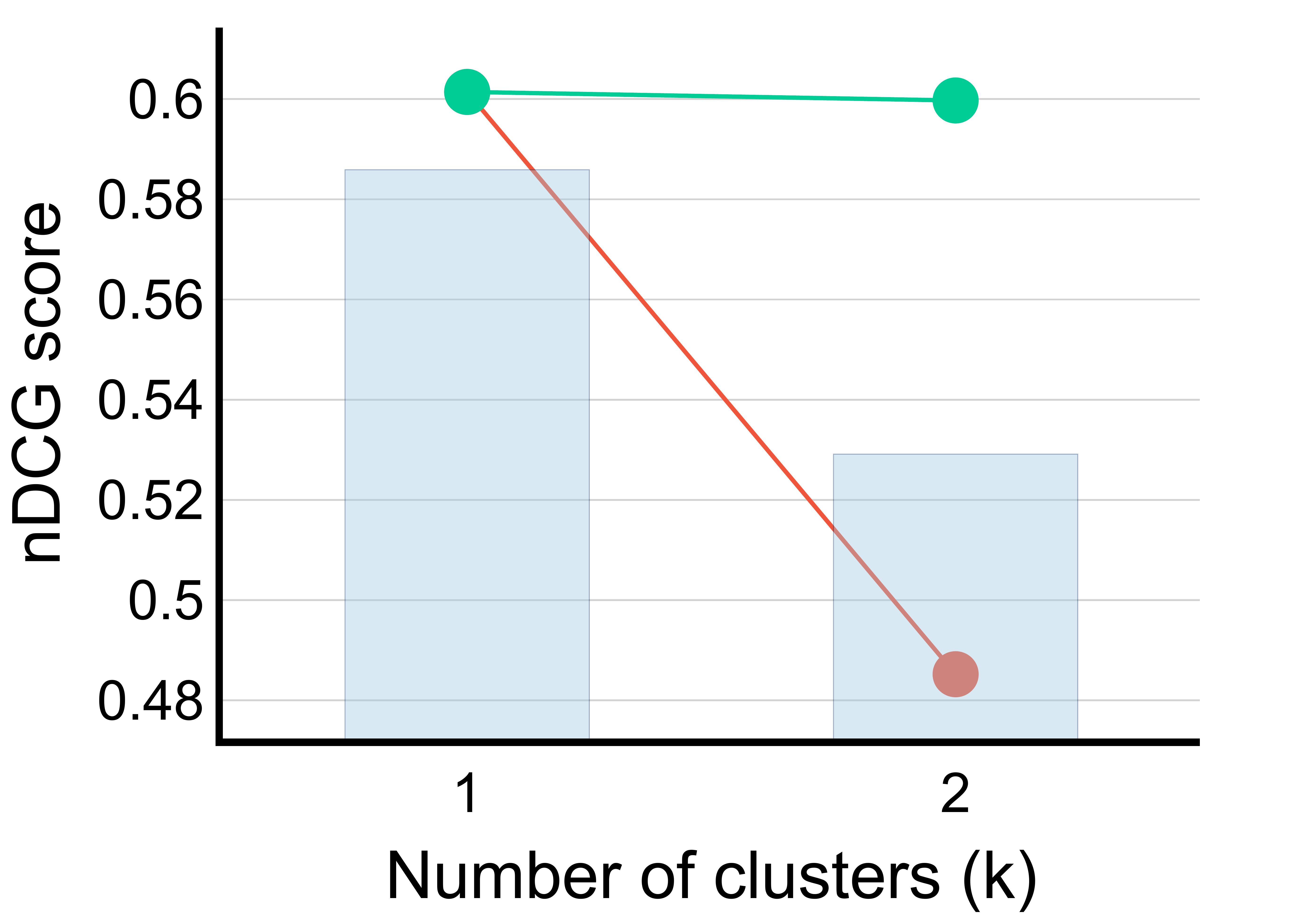} 
 & \includegraphics[scale=0.07,type=pdf,ext=.pdf,read=.pdf]{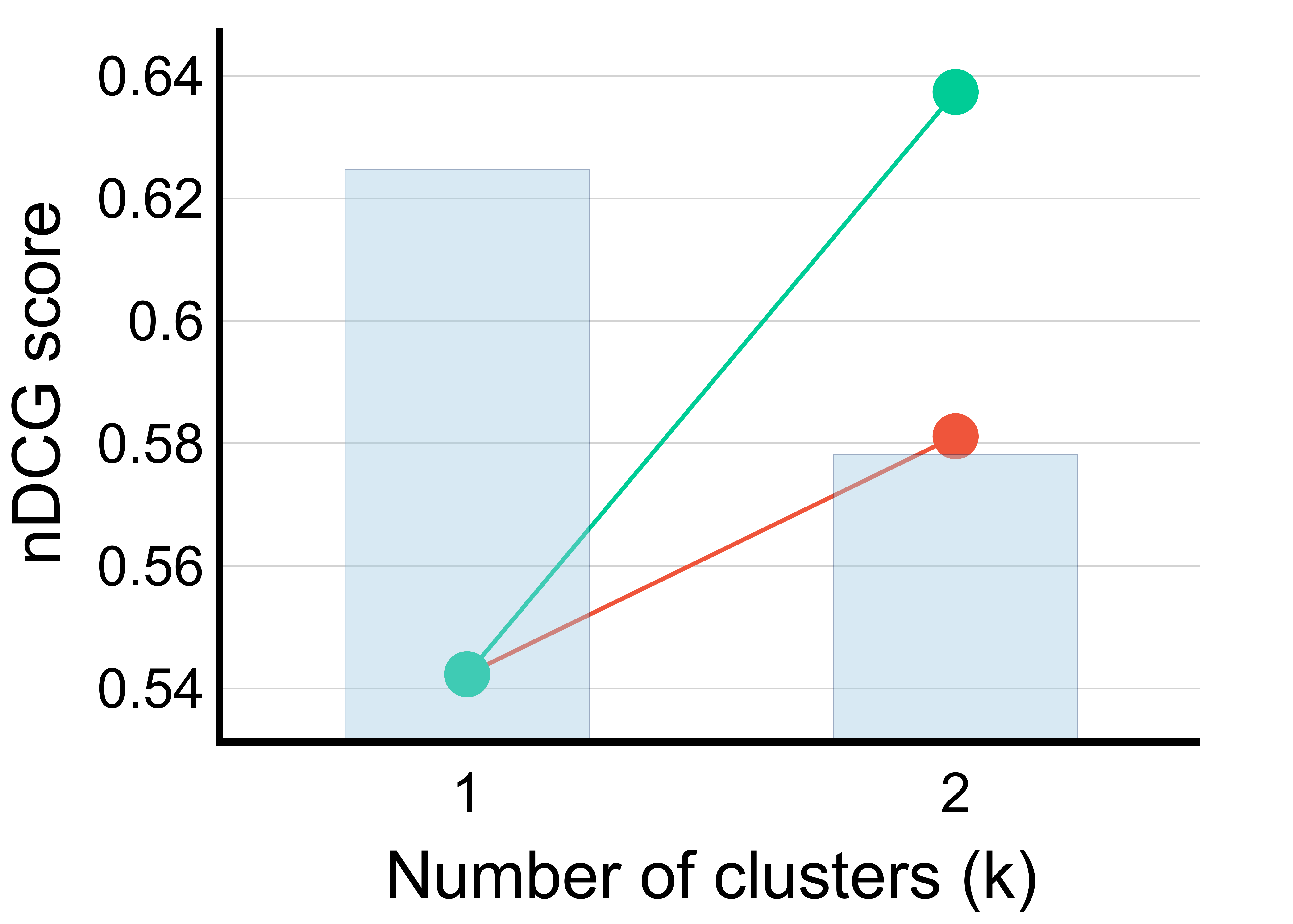} 

  & \includegraphics[scale=0.07,type=pdf,ext=.pdf,read=.pdf]{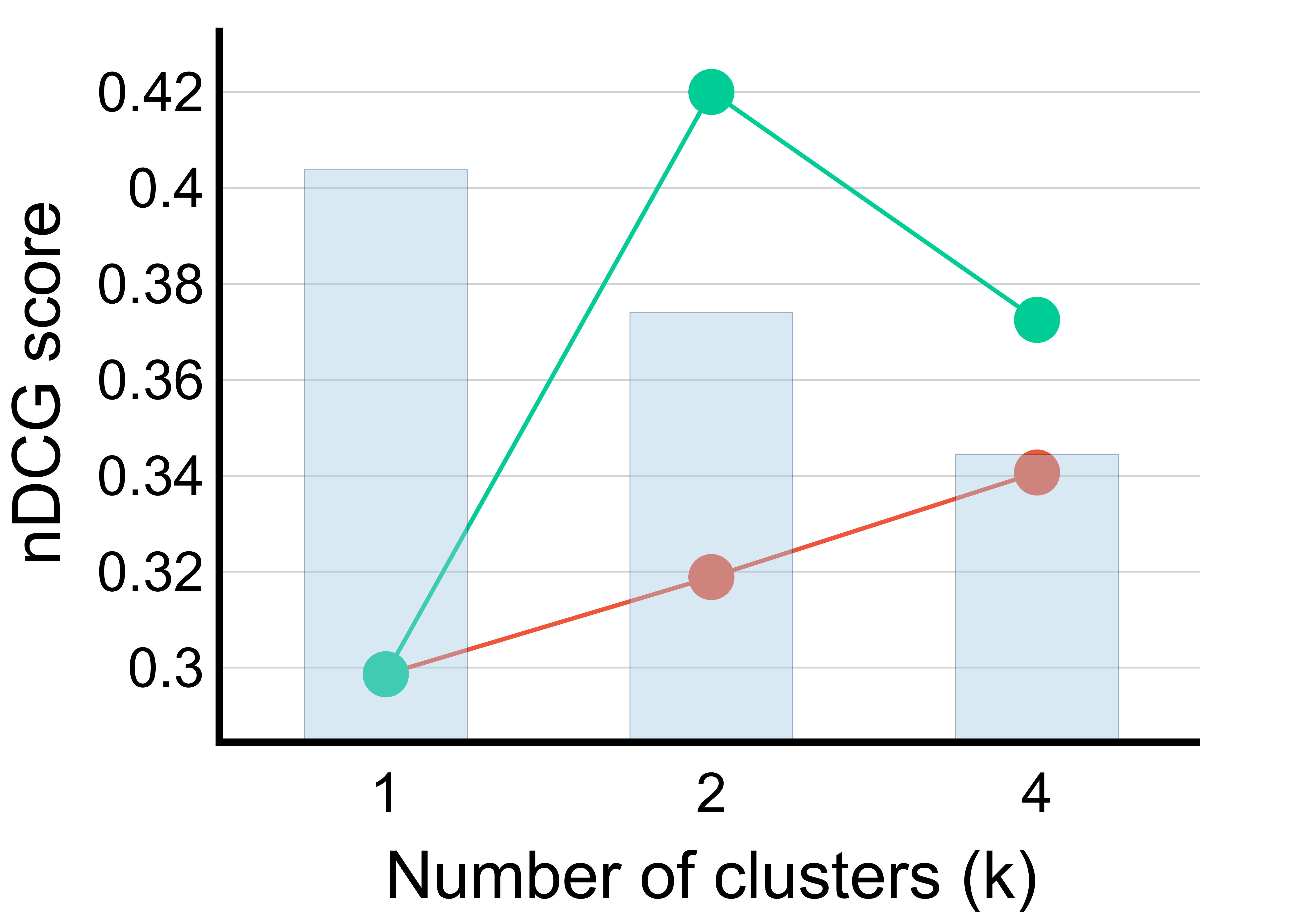} 
& \includegraphics[scale=0.07,type=pdf,ext=.pdf,read=.pdf]{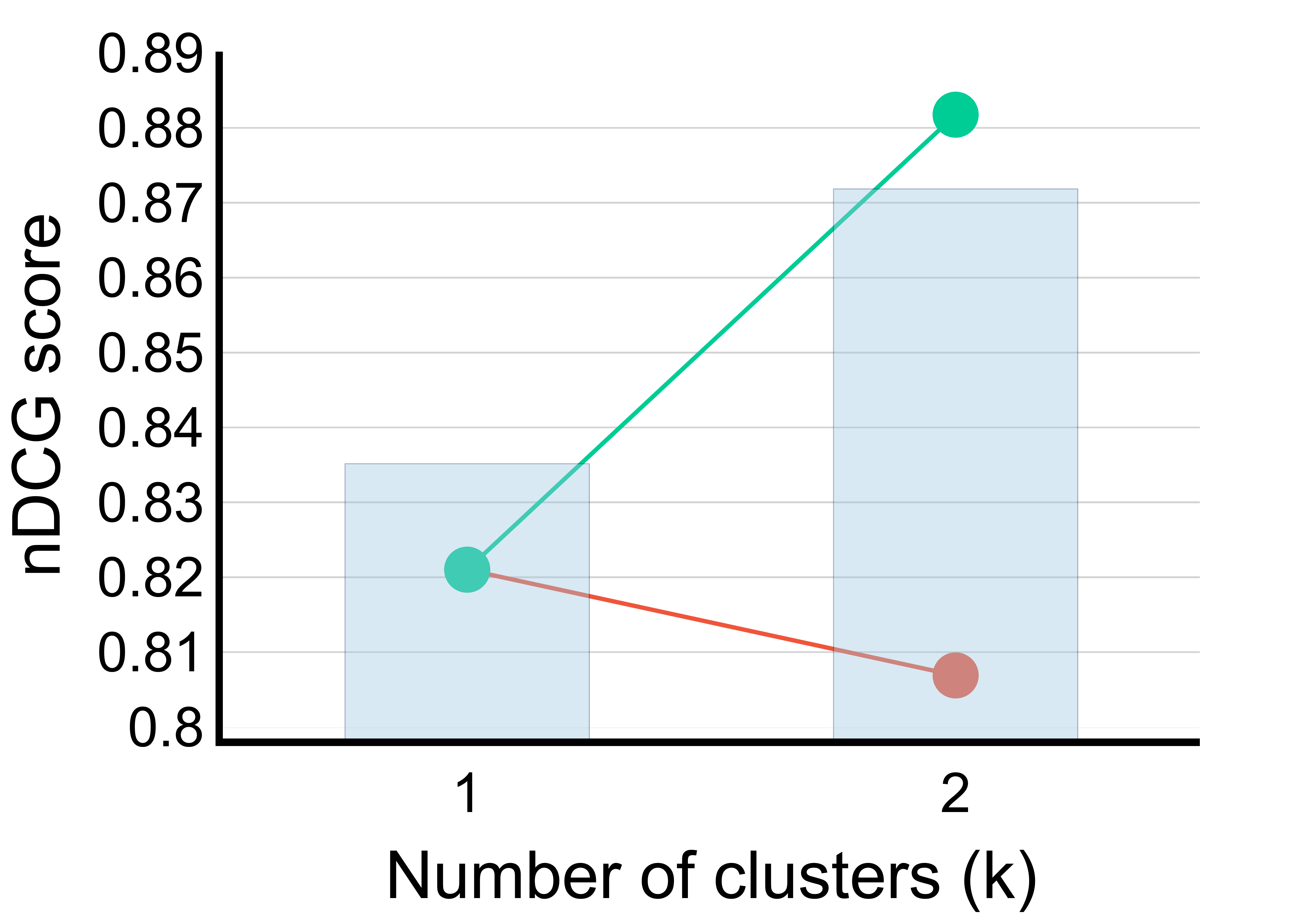} 
  \includegraphics[scale=0.1]{fig/legend_varying-k.pdf}  
\end{tabular}
\caption{Compressed size and AUC or nDCG score vs. $K$ for
  APT scenario 1}\label{fig:process-results-s1}
\end{figure*}

To assess Q2, we ran \AVCstar, \Krimpstar and \CompreXstar on all of
the datasets, for different values of $k$ (1, 2, 4, 8, 16, 20 for
\AVCstar and \Krimpstar, but only 1,2,4 for \CompreXstar because the
running time was prohibitive for higher values of $k$).
Figures~\ref{fig:bank-probe}-\ref{fig:process-results-s1} show
selected results from these experiments, plotting relative compressed
sizes (bars) and maximum and median AUC or nDCG scores (green and red
lines) against $k$ values.  In this experiment we ran 10 trials of
each algorithm for a fixed value of $k$ and again report medians of
AUC or nDCG scores for the 10 runs.  For illustration, we also show
the maximum AUC or nDCG score achieved in each batch of 10 runs as
well as the median; however, in an unsupervised setting we have no way
of knowing in advance which of several runs will produce this optimal
result.

\begin{figure*}[tb]
  \begin{tabular}{ccc}
Public & APT Scenario 1 & APT Scenario 2\\
\includegraphics[scale=0.09]{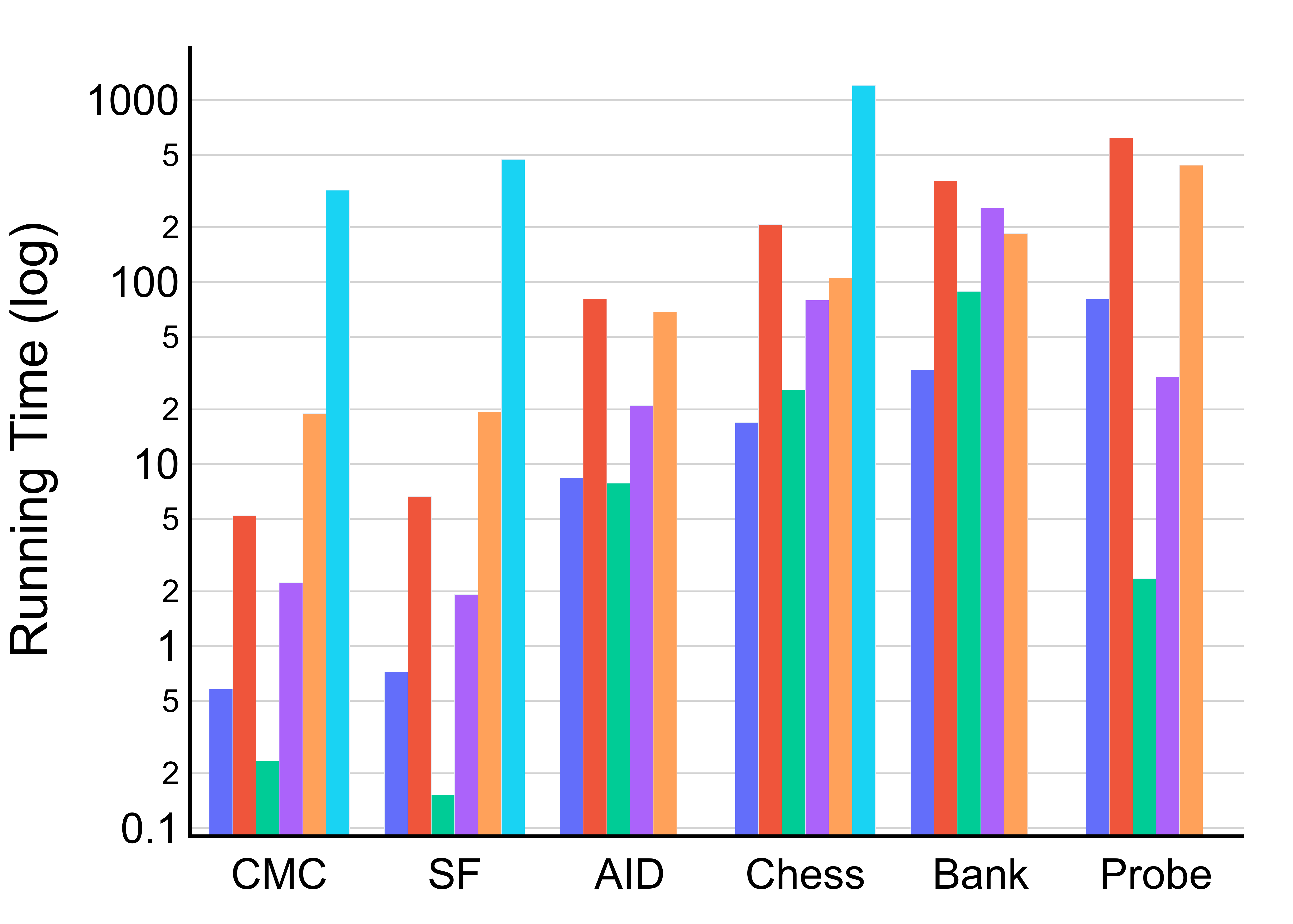}
&
\includegraphics[scale=0.09,type=pdf,ext=.pdf,read=.pdf]{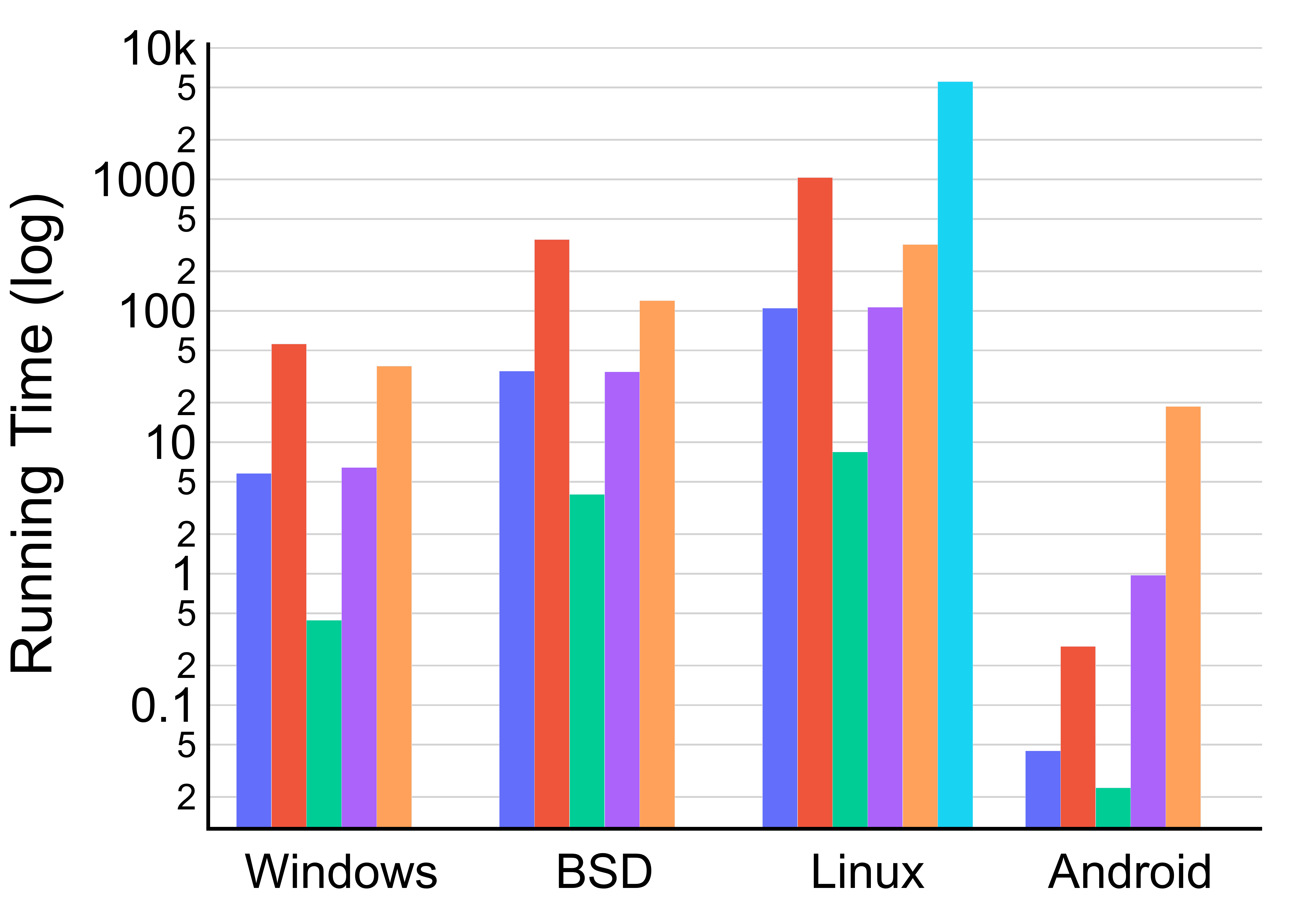}
&
\includegraphics[scale=0.09,type=pdf,ext=.pdf,read=.pdf]{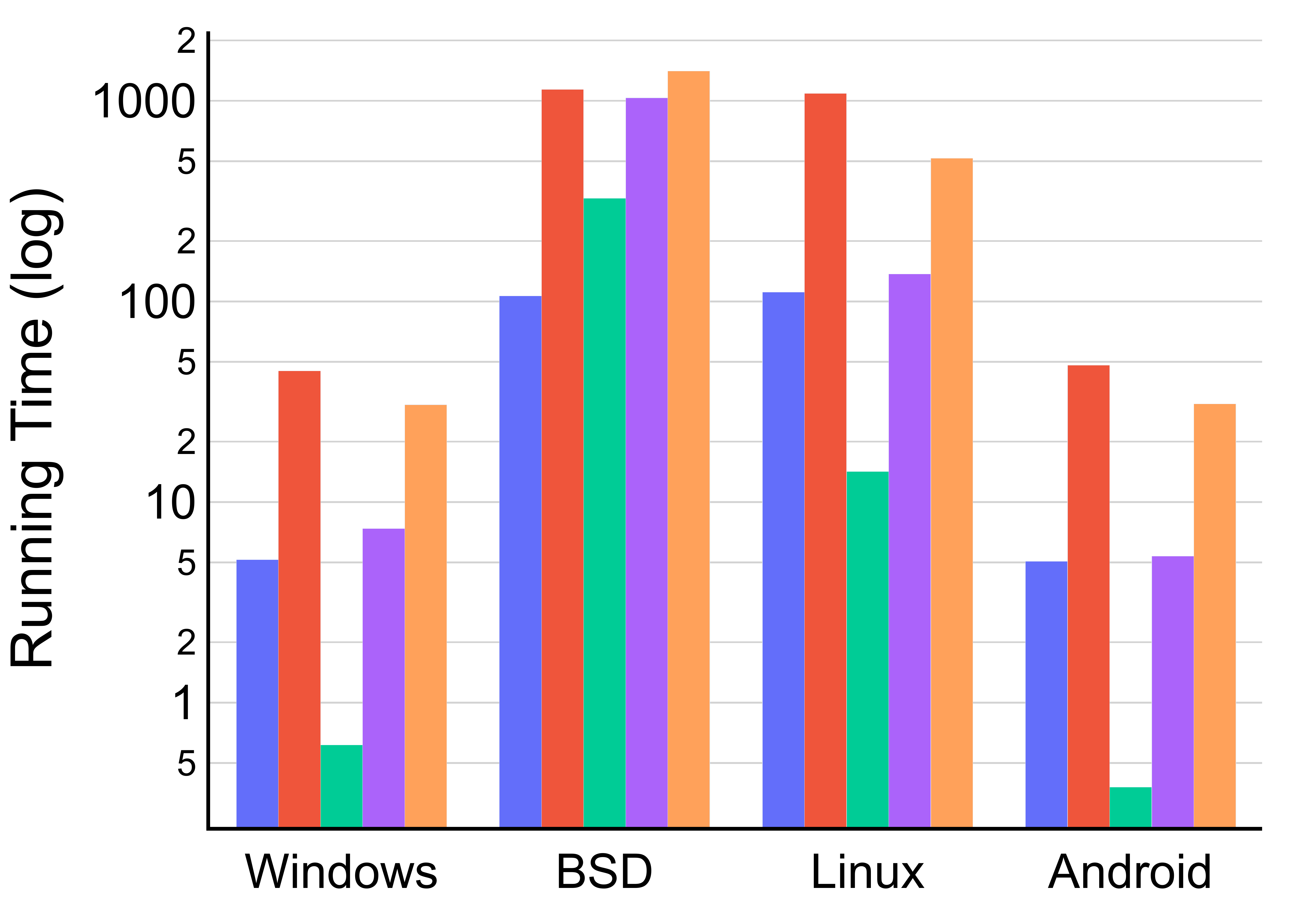}
\includegraphics[scale=0.1]{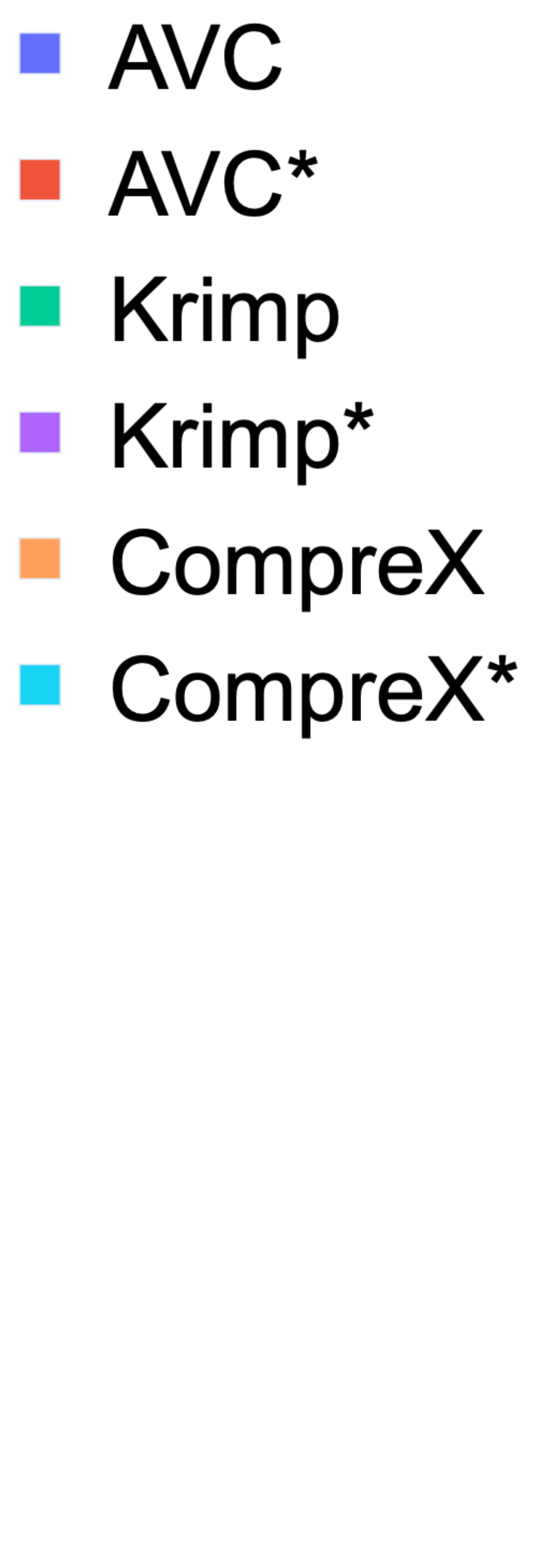}
  \end{tabular}
\caption{Runtime in seconds for general, APT scenario 1 and APT
  scenario 2 datasets}\label{fig:runtime}
\end{figure*}

Finally to assess Q3, we report the average running times (again
across 10 runs) of each algorithm on each dataset.  These are shown in
Figure~\ref{fig:runtime}.  The reported running times for AVC, OC3,
CompreX, \AVCstar and \Krimpstar are for 10
runs of the full algorithm, considering all $k$-values up to 20, with
early stopping if a local minimum is identified early.  For
\CompreXstar we report only a few examples for $k$-values up to 2 or 4
because each run takes over a minute even for small datasets.
In
interpreting these results it is important to keep in mind that each
base algorithm was implemented in a different programming language:
AVC in Python (using libraries such as numpy for efficient matrix
manipulations), Krimp in C++, and CompreX in Matlab.  Moreover, the
wrapper Python code for \AVCstar, \Krimpstar, and \CompreXstar may
contribute to higher overhead for these algorithms (for example due to
repeated process startup costs) compared to a single-language
implementation.  Nevertheless, these results allow at least a coarse
qualitative comparison among the different techniques.

\subsection{Discussion}

\paragraph{Effectiveness of \AVCstar compared to OC3 and CompreX} 
The results in Table~\ref{tab:q1results} show that generally, AVC is
not competitive with Krimp/OC3 or CompreX, with the interesting
exception of the Probe dataset.  However, when we consider mixtures of
AVC models in \AVCstar, anomaly detection performance  increases
significantly, leading to the best overall results in 10 cases,
compared to 13 for CompreX and six for Krimp/OC3.  (When there is a
tie, we give credit to both techniques.)  Thus, despite its simplicity, the
\AVCstar algorithm illustrates that a simple MDL-based compression
model together with MDL-based clustering yields a competitive anomaly
detection technique.

\paragraph{Effect of clustering on anomaly detection}
For \AVCstar, increasing the number of clusters typically improves
compression performance.  On the other hand, for \Krimpstar, the best
compression often results from $K=1$ and typically fewer clusters are
found.  
For the generic datasets, increasing $K$ does not always
translate to improved anomaly detection performance, even if it
improves compressed size.  As
representative examples, consider the Bank and Probe
datasets in Figure~\ref{fig:bank-probe}.  Both are compressed more effectively
by \AVCstar for $k = 16$ or $20$ while the best $k$ value for
\Krimpstar is 1.   Increasing $K$ results in improved
median anomaly detection scores by \AVCstar while for the other
situations (Bank using \Krimpstar and Probe using either algorithm)
increasing $K$ leads to no improvement.  The counterintuitive results
for Probe could result, for example, if there are several clusters and
all of the anomalies are close to one large cluster but far from
representative of the dataset as a whole.  

We also consider the results obtained for the APT security datasets.
We show the results for Scenario 1 in Figure~\ref{fig:process-results-s1}, since the Scenario 2 results are
similar, and because \AVCstar was more effective than Krimp or CompreX
on the Linux and Android datasets for Scenario 1.  The Windows and BSD
datasets display a clear trend: for \AVCstar, decreasing compressed
size corresponds to improved anomaly detection.  \Krimpstar and
\CompreXstar show improvements in some cases but these are more equivocal.
For Linux scenario
1, \AVCstar's anomaly detection performance again increases with
larger $K$, corresponding to a decrease in compressed size.  This is
also the case for \CompreXstar, while \Krimpstar does obtain lower
compressed size with 4--8 clusters but the increase in anomaly
detection performance is not as significant, with median AUC score
nearly unchanged and nDCG score increasing from 0.34 to 0.41.  On the
other hand, for the Android dataset (the smallest of the APT
datasets), \AVCstar obtains only small improvements in compressed size
for $K=2$ compared to $K=1$, with some associated improvement in AUC
and nDCG scores, while both \Krimpstar and \CompreXstar obtain minimal
compressed size at $K=1$.  Again, the general trend is that \Krimpstar
and \CompreXstar find fewer clusters.

Overall, these results show that for \AVCstar, there is usually a
clear relationship between finding more clusters, decreased compressed
size, and improved anomaly detection performance.  For \Krimpstar and
\CompreXstar, there are improvements but they are not as pronounced, possibly
because the basic Krimp and CompreX models are already sufficiently capable of
adapting to heterogeneity, so that fewer clusters are needed to find a
good model of the data.  

\paragraph{Running time}
Figure~\ref{fig:runtime} shows the average running time for each
technique on each dataset (with the exception of \CompreXstar which we
only report for a few datasets).  Each run of \AVCstar or \Krimpstar
potentially considers all $k$-values up to 20 (and up to 2 or 4 in the
case of \CompreXstar).  A clear trend is that the clustering version
of each algorithm takes several times longer than the basic algorithm;
in the case of \CompreXstar the slowdown can be as much as a factor of
10, making CompreX and \CompreXstar less competitive in terms of
performance.  In general, we find that CompreX runs slower than OC3,
contrary to the results reported by Akoglu et al.~\cite{comprex}, but
this may be due to differences in maturity of the version of Krimp
used.  Generally, Krimp is faster than AVC, which is faster than
CompreX, and likewise for the clustering variants \Krimpstar, \AVCstar
and \CompreXstar respectively.  However, it should be recalled that
each algorithm is implemented in a different language (Python, C++ or
Matlab) so this may explain some of the differences.  In particular,
further optimization or
parallelization of \AVCstar should be effective, because it is
algorithmically much simpler than Krimp and there are many unexploited
opportunities for parallel processing.

\section{Related Work}\label{sec:related}

Mixture models and clustering have been used for anomaly detection in
numerical and mixed data; for example, the SmartSifter
algorithm~\cite{smartsifter} considers mixed categorical and numerical
data, and induces a different mixture model of numerical data for each
combination of categorical attributes. SmartSifter also uses an
MDL-like logarithmic anomaly score. However, SmartSifter's running
time grows exponentially in the number of categorical attributes, and
in practice, scales with the number of combinations of attributes
actually present in the data, which makes it unsuitable to datasets
with large numbers of categorical attributes. Another approach due to
Bouguessa aggregates the results of several anomaly detectors and fits
a mixture model to identify anomalous clusters~\cite{bouguessa14jait}.

Clustering techniques have been widely considered for numerical
(non-categorical) data, while clustering for categorical data (which
usually lacks natural metrics) has received much less study; the main
approaches considered so far include EM-style algorithms for latent
class inference or fitting discrete mixture models~\cite{bishop06};
k-modes which performs clustering with respect to a dissimilarity
metric~\cite{huang97dmkd}; and MDL-based
clustering~\cite{kontkanen06mdl,identifying-the-components}.

The MDL-based clustering approach we adopted is inspired by and
similar to those of Kontkanen et al.~\cite{kontkanen06mdl} and van
Leeuwen et al.~\cite{identifying-the-components}, but differs from
both in that we consider clustering based on any MDL-based technique,
whereas Kontkanen et al. consider a minimax optimal encoding of
mixtures of discrete distributions, and van Leeuwen et al. consider
Krimp as the base compressor but their approach does not take into
account the cost of encoding the inferred classes.  The latter do
observe that their k-means-style algorithm could be used with other
compressors. Moreover, neither approach has previously been considered
as a basis for anomaly detection, whereas previous work on MDL-based
anomaly detection has not considered clustering or mixture model
fitting. Our work shows that this approach can improve anomaly
detection for a variety of MDL techniques, resulting in a new
competitive algorithm \AVCstar and in some cases improving on the
performance of Krimp/OC3 and CompreX reported in previous work.

Besides Krimp/OC3 and CompreX, another MDL-based approach to anomaly
detection is the UPC algorithm of Bertens et al.~\cite{upc}, which
uses a Krimp-style compression algorithm and then looks for objects
with unusual \emph{combinations} of features.  The anomaly scores are
not directly based on compressed sizes.  To the best of our knowledge
the only previous work applying clustering to categorical anomaly
detection is the ROAD algorithm~\cite{suri14ijhis}, and a variant
based on rough sets called Rough-ROAD~\cite{suri16nc}. Both algorithms
perform clustering based on an ad-hoc metric on categorical data, and
neither is based on MDL. ROAD was shown to have better performance
than AVF~\cite{avf} but was not compared with other anomaly detection
algorithms such as Krimp/OC3 or CompreX, while Rough-ROAD was compared
only with ROAD.

This work has been motivated by the observation that according to the
minimum description length principle, the best-fitting model for some
data is the one that minimizes the communication cost of the data,
including the cost of describing the chosen model. However, to apply
MDL requires deciding on a class of models and encoding scheme for
them.  Moreover, allowing more complex models increases the cost and
algorithmic difficulty of searching and fitting the best one. In the
limit, we could consider arbitrary compression algorithms as predictive models
but then finding the optimal one would be undecidable.  Allowing
mixture models and fitting them using clustering is one strategy for
enriching the model space, which allows for the possibility of
improved compression (that is, better prediction), while remaining
relatively algorithmically straightforward. The idea of CompreX, that
is, partitioning the columns of a dataset to enhance compression, is
another approach. It would be interesting to explore other points on
the tradeoff curve between model (and search space) complexity and
compression effectiveness.

The public datasets for anomaly detection are drawn from those
collected by Pang et al.~\cite{pang16icdm,pang16ijcai} in their work
on feature selection and anomaly detection. Their work demonstrates
how unsupervised feature selection could be used to improve
categorical anomaly
detection, including for CompreX; this is an orthogonal direction for
improving anomaly detection effectiveness and performance, and it may
be interesting to study whether it can be combined with
clustering/mixture modeling.

Our main motivating application has been data gathered from security
exercises in a recent DARPA program. These datasets are extracted from
a much richer graph dataset in which operating system processes,
files, and other resources are represented as nodes, and relationships
between them as edges. Anomaly detection and intrusion detection
techniques have been developed and applied to these datasets
and studied by a number of 
papers~\cite{berrada20fgcs,berrada19tapp,han2020unicorn,milajerdi19holmes,milajerdi19poirot,siddiqui18kdd},
but there is as yet no commonly accepted public dataset of ground
truth annotations for evaluating different techniques; we used those
developed by the authors of~\cite{berrada20fgcs,berrada19tapp} for
their binary feature datasets. Comparing or combining our work with
the above results obtained by others would be a valuable exercise.

\section{Conclusions}\label{sec:concl}

Anomaly detection over categorical data is challenging but has received
comparatively less attention than for continuous or numerical data.
The previous state of the art is to search for patterns in the data
that aid compression, for example using itemset mining or partitioning
the columns into mutually informative subsets, and then use the
codelength of each record as its anomaly score; according to the
MDL principle, the best model is the one that compresses the data
best, so once such a model is found, the anomalies are the records
compressed poorly by the model.

We observe that in heterogeneous datasets containing mixtures of
distinct kinds of records, existing techniques may miss opportunities
to improve compression, and hence allowing \emph{mixtures} of models
may improve compression and lead to better-fitting models according to
the MDL principle. Though the idea of fitting mixtures of models to
data is not new, and has been used as a basis for MDL-based clustering
techniques, we propose the use of mixture models and clustering
algorithms to improve the performance of MDL-based anomaly detection.
We illustrated this general strategy using three MDL-based anomaly
detection techniques as the components of mixture models: simple
discrete AVC models, the Krimp algorithm underlying
OC3, and the CompreX algorithm.

Our results show that, in many cases, using a k-means-style algorithm
finds opportunities for improving compression compared to using a
single model. Moreover, we also show that mixture modeling can improve
the anomaly detection performance of existing algorithms such as
Krimp/OC3 and CompreX, and even that mixtures of simple models provide
competitive anomaly detection performance compared to unmixed Krimp or
OC3 models. On the other hand, performing iterative clustering to fit
a mixture model is more expensive computationally, and our choice of
randomized initialization of the clusters may leave room for further
improvement since it is well known that randomly initializing clusters
is not always the best strategy. Finally, we considered only the
case of batch, nonadaptive processing and it would also be interesting
to consider streaming, adaptive anomaly detection techniques based on
incremental clustering.


\bibliography{paper}
\bibliographystyle{abbrv}

\end{document}